\documentclass{aa}  

\usepackage{graphicx}
\usepackage[varg]{txfonts}
\usepackage{graphicx}	
\usepackage{amsmath}	
\usepackage{xcolor}
\usepackage{multirow}
\usepackage{tablefootnote}
\usepackage[T1]{fontenc}
\usepackage{lscape}
\usepackage{booktabs}
\usepackage[toc]{appendix}
\usepackage{lipsum}
\usepackage{stfloats}
\usepackage{calc}
\usepackage{subcaption}
\usepackage{orcidlink}
\usepackage{footmisc}

\newcommand{\orcid}[1]{\href{https://orcid.org/#1}{\includegraphics[width=9pt]{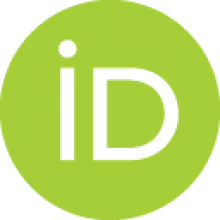}}}
\usepackage{natbib,twoopt}
\hypersetup{
    colorlinks = true,
    linkcolor = {blue},
    citecolor = {blue}
}
\usepackage{orcidlink}
\let\orcid\orcidlink

\bibpunct{(}{)}{;}{a}{}{,} 
\makeatletter
\newcommandtwoopt{\citeads}[3][][]{\href{http://adsabs.harvard.edu/abs/#3}%
{\def\hyper@linkstart##1##2{}%
\let\hyper@linkend\@empty\citealp[#1][#2]{#3}}}

\newcommandtwoopt{\citepads}[3][][]{\href{http://adsabs.harvard.edu/abs/#3}%
{\def\hyper@linkstart##1##2{}%
\let\hyper@linkend\@empty\citep[#1][#2]{#3}}}

\newcommandtwoopt{\citetads}[3][][]{\href{http://adsabs.harvard.edu/abs/#3}%
{\def\hyper@linkstart##1##2{}%
\let\hyper@linkend\@empty\citet[#1][#2]{#3}}}

\newcommandtwoopt{\citeyearads}[3][][]%
{\href{http://adsabs.harvard.edu/abs/#3}
{\def\hyper@linkstart##1##2{}%
\let\hyper@linkend\@empty\citeyear[#1][#2]{#3}}}

\makeatother

\begin{document} 

\titlerunning{AGN outflows in dwarf galaxies}
\authorrunning{Arjona-Gálvez E. et. al}

\title{AGN-driven outflows in dwarf galaxies from cosmological simulations:}

\subtitle{Internal properties and observational signatures}

\setcounter{footnote}{1}

\author{Elena Arjona-Gálvez
\orcidlink{0000-0002-0462-7519}\inst{1,2}\thanks{\email{elenarjonagalvez@gmail.com}}%
, Arianna Di Cintio \orcid{0000-0002-9856-1943} \inst{2,1}, Robert J. J. Grand \orcid{0000-0001-9667-1340} \inst{3}, Laura V. Sales \orcid{0000-0002-3790-720X}\inst{4}, Gabriela Canalizo \orcid{0000-0003-4693-6157}\inst{4}, Teresa Matamoro Zatarain \orcid{0009-0007-2976-293X}\inst{5} and Aswin P. Vijayan \orcid{0000-0002-1905-4194}\inst{6}}

\institute{Instituto de Astrofísica de Canarias, Calle Vía Láctea s/n, E-38206 La Laguna, Tenerife, Spain
\and 
Universidad de La Laguna, Avda. Astrofísico Fco. Sánchez, E-38205 La Laguna, Tenerife, Spain
\and
Astrophysics Research Institute, Liverpool John Moores University, 146 Brownlow Hill, Liverpool, L3 5RF, U,
\and
Department of Physics and Astronomy, University of California, Riverside, CA 92507, USA,
\and
School of Physics, HH Wills Physics Laboratory, University of Bristol, Tyndall Avenue, Bristol BS8 1TL, UK,
\and
Astronomy Centre, University of Sussex, Falmer, Brighton BN1 9QH, UK
}

\date{Received XXX, XXXX; accepted YYY, YYY}

 
\abstract
{}
{While active galactic nucleus (AGN) feedback is a key driver of massive galaxy evolution, its physical properties and observational signatures in the dwarf regime remain poorly understood. We investigate the impact of AGN-driven outflows on the interstellar medium (ISM) of dwarf galaxies and assess whether these events can be robustly identified through emission-line diagnostics.
}
{We analysed a high-resolution cosmological magneto-hydrodynamical zoom-in simulation from the \texttt{AURIGA} project. We focused on a dwarf galaxy with $\rm M_{\star}$$\sim$$10^{9.7}\,\rm M_\odot$ hosting a massive black hole (BH) of $\rm M_{BH}$$\sim$$10^{7}\,\rm M_\odot$. We identified individual outflow episodes via pressure peaks in the gas surrounding the central BH, tracked the thermodynamic and kinematic history of such gas, and computed synthetic, spatially resolved nebular emission using photoionisation models to construct BPT diagnostic diagrams.
}
{We show that AGN activity in this mass regime produces compact, over-pressurised central bubbles reaching extreme temperatures, T$>$$10^{6}$\,K. These structures accelerate the multiphase ISM to velocities up to $\sim$600$\,\rm km\,s^{-1}$, significantly exceeding those driven by stellar feedback only: the outflowing material does not escape the halo, but instead decelerates and redistributes within $\sim$10\,kpc from the galaxy center. Synthetic emission-line modelling reveals clear, time-dependent signatures of such AGN-driven feedback. Over its life cycle, the simulated AGN-hosting galaxy traces the locus of observed dwarf AGNs, and migrates from the star-forming sequence in the BPT diagrams through the composite region and into the AGN regime, highlighting a self-regulation mechanism in which the BH accretes its fuel supply, progressively moving towards the low-ionization nuclear region.}
{Our results suggest that AGN-driven outflows in dwarf galaxies primarily regulate the central ISM through episodic heating and rapid gas recycling, rather than large-scale gas ejection. These processes generate observable spectroscopic signatures, offering a promising avenue for identifying AGN feedback in low-mass galaxies.}

\keywords{method: numerical -- galaxies: active -- galaxies: dwarfs -- galaxies: evolution -- galaxies: formation -- galaxies: star formation}

\maketitle

\section{Introduction}\label{sec:INTRO}

Feedback from active galactic nuclei (AGN) is widely recognised as a key driver of galaxy evolution \citep{Silk&Rees1998,King2003}. The immense energy released as gas accretes onto massive black holes (BH) can regulate star formation (SF) by heating and expelling the interstellar medium (ISM), as well as enrich the circumgalactic medium (CGM), and establish scaling relations, such as the correlation between BH mass and stellar velocity dispersion (M$_{\rm BH}$–$\sigma_\star$; \citep{Kormendy2013}). In massive galaxies, these processes are essential for reproducing the observed quenching of SF and the bright-end cutoff of the galaxy luminosity function \citep[see e.g.][and references therein]{Fabian2012,Harrison2018}.

While the impact of AGN in high-mass systems is well established, their role in low-mass galaxies remains far less clear. In dwarf galaxies (M$_\star \lesssim 10^9$–10$^{10}$ M$_\odot$), with their shallow gravitational potentials and relatively simple structures, the galaxy evolution models have long attributed the regulation of SF primarily to stellar feedback, such as supernovae and radiation from young stars \citep{Dekel&Silk1986,MacLow&Ferrara1999,Hopkins2014}. Powerful supernova-driven winds are observed in starbursting dwarfs like M82 \citep{Martin1998,Strickland&Stevens2000}, and cosmological simulations routinely invoke strong stellar feedback to explain the low SF efficiencies of these systems \citep{Governato2010}.

However, mounting evidence challenges the view that AGN are negligible in dwarfs. Large surveys have revealed a surprisingly high incidence of AGN in nearby low-mass galaxies \citep{Reines2013,Mezcua2018,Reines2022}, implying that BHs can grow even in small halos. Spatially resolved spectroscopy has begun to uncover direct signatures of AGN-driven outflows in these systems. \cite{ManzanoKing2019} detected high-velocity [O\,III] winds in SDSS-selected dwarf AGN, with velocities exceeding the escape speeds of their halos. Follow-up studies with Keck/KCWI confirmed extended ionised outflows in nearby dwarfs \citep{Liu2020}, while MaNGA observations revealed broad emission-line components in low-mass AGN hosts with kinetic powers inconsistent with purely stellar origins \citep{RodriguezMorales2025, Mezcua2024}. Complementary ultraviolet absorption-line studies further highlight the multiphase nature of these winds, detecting highly ionised, fast-moving gas in dwarf systems \citep{Liu2024}.

Beyond their detection, a key open question concerns the physical impact of AGN-driven outflows on their host galaxies. Outflows are often invoked as mechanisms that can remove large quantities of gas from galaxy spheroids, potentially shutting down SF. Yet theoretical work has emphasised that they can also compress dense gas clouds, enhancing fragmentation and possibly triggering SF under certain conditions \citep{SilkNusser2010,ZubovasBourne2017,Zubovas2023}. This dual behaviour, where outflows can simultaneously suppress and locally enhance SF, is seen not only in analytic frameworks but also in cosmological simulations, where quasar winds depress overall SF rates while creating pockets of locally increased SF through gas compression \citep{MercedesFeliz2023}. Observational evidence for this complex interplay remains limited but suggestive; for example, spatially resolved observations of quasars have uncovered clear instances of simultaneous positive and negative feedback \citep{Bessiere2022}. Additional evidence for positive feedback includes cases of jet-induced SF \citep{Crockett2012}, detections of SF occurring within AGN-driven outflows in more massive systems \citep{Maiolino2017}, as well as potential triggers in the low-mass regime like Henize 2-10 \citep{Schutte&Reines2022}. Disentangling these competing mechanisms is further complicated by the short duty cycles of AGN activity \citep{Schawinski2015}, making induced SF episodes difficult to isolate \citep{KingNixon2015}. It therefore remains unclear whether gas removal or SF triggering dominates over galaxy lifetimes, particularly in the low-mass regime.

On the theoretical side, the impact of AGN on dwarfs is still debated. Some analytic and idealised models suggest that AGN feedback could, in principle, be more efficient than stellar feedback at clearing gas in low-mass halos \citep{Silk2017,Dashyan2018}. Others argue that supernova-driven turbulence prevents sustained BH growth by inhibiting the infall of dense, cold gas, rendering AGN energetically subdominant \citep{Dubois2015,Habouzit2017,Trebitsch2018,Koudmani2019,Koudmani2021}. Conversely, recent high-resolution simulations exploring different accretion and feedback prescriptions show that BHs can grow efficiently and drive impactful outflow in the dwarf regime \citep{Sharma2020,Koudmani2022,Wellons2023}. In our previous work, we used high-resolution cosmological simulations to show that even low-luminosity AGN can measurably affect dwarf galaxies: accreting BHs heat and displace central gas reservoirs, reduce SF, and in some cases lower central dark matter densities by up to $\sim$65\% without fully depleting the gas content \citep{ArjonaGalvez2024}. 

A related challenge concerns how to quantify the energetic relevance of observed outflows and connect them to theoretical predictions. The ratio between outflow kinetic power and AGN bolometric luminosity, often referred to as the kinetic coupling efficiency ($\epsilon_f$), is widely used to compare observations with simulations. However, such comparisons are not straightforward. In many cosmological models, efficiencies of $\sim$0.5\%-5\% are typically assumed to reproduce local scaling relations \citep[e.g.][]{DiMatteo2005,Springel2005} yet only a fraction of this injected energy effectively translates into large-scale kinetic motion. Observational estimates of kinetic power are highly sensitive to assumptions regarding outflow geometry and the multi-phase nature of the gas \citep[see][for a comprehensive review]{HarrisonRamosAlmeida2024}. Furthermore, the presence of compact jets can significantly accelerate gas even in radiatively efficient AGN, complicating interpretations based solely on bolometric luminosities. These uncertainties highlight the need for more physically grounded approaches that can directly bridge the gap between simulated outflow physics and observational diagnostics.

Despite recent progress, a key gap remains between simulations and observations in the dwarf regime. While simulations can track the physical properties of AGN-driven outflows, observational identification relies primarily on emission-line diagnostics and kinematic signatures. Standard emission-line classification schemes such as BPT and VO87 diagrams \citep{Baldwin1981, Veilleux1987} are essential tools in large spectroscopic surveys. However, it remains unclear how AGN-driven outflows in dwarf galaxies populate these diagnostic spaces. The typically low metallicity and high specific SF rates of dwarf galaxies can shift AGN signatures towards the SF or composite regions of the diagnostic diagram, potentially hiding a significant population of active BHs \citep{Trump2015,Cann2019}. This highlights the necessity of using theoretical models to predict the expected line ratios of AGN systems, specifically in the low-mass regime, where traditional thresholds may not apply. In this work, we aim to bridge this gap by comparing our simulations to the limited observational samples of local dwarf galaxies currently available \citep[e.g.][]{Moran2014,Reines2013,ManzanoKing2019,Liu2020}.

By leveraging high-resolution cosmological magneto-hydrodynamical zoom-in simulations from the \texttt{AURIGA} project \citep{Grand2017,Grand2024}, previously analysed in \cite{ArjonaGalvez2024}, we track the life cycle of individual AGN-driven outflows. Our analysis provides a time-resolved view of how energy injection from a central BH reconfigures the gas structure and kinematics of its host. We compute spatially resolved nebular emission using \texttt{Synthesizer} \citep{Lovell2025,Roper2025}, which self-consistently translates the intrinsic BH parameters into predicted optical emission-line ratios (Vijayan et al. in prep). Furthermore, by tracing the evolution of emission-line ratios across different accretion states of the BH, we investigate how AGN variability and outflow episodes shape the spectroscopic classification of dwarf galaxies over time. This approach provides a physically grounded framework to interpret AGN feedback in the low-mass regime and offers testable predictions for current and upcoming spectroscopic surveys. The paper is organised as follows: Section \ref{sec:methodology} describes the simulations and emission modelling. Sections \ref{sec:outflowproperties} and \ref{sec:kinematics} present the physical properties of the outflows, while in section \ref{sec:obscomparison}, we analyse their synthetic observational signatures. We discuss our results in section \ref{sec:conclusions}.

\section{Methodology}\label{sec:methodology}
\subsection{Simulation suite}

In this work, we use the subset of cosmological magneto-hydrodynamical zoom-in simulations from the \texttt{AURIGA} project \citep{Grand2017,Grand2024} used in \cite{ArjonaGalvez2024}.  The simulations were run with the moving-mesh code \texttt{AREPO} \citep{Springel2010,Pakmor2016}. The sample is composed of 12 isolated haloes, with a $z$=$0$ mass between $5\times10^{10}\,\rm M_{\odot}$ and 5$\times$10$^{11}\,\rm M_{\odot}$, selected from the DM-only \texttt{EAGLE} simulation, in a co-moving box with a side length of 67.77$\,h^{-1}\,\rm cMpc$ (L100N1504), as introduced in \cite{Schaye2015}. The cosmological parameters adopted for this sample are given in \cite{Planck2014}: $\Omega _m$=$0.307$, $\Omega _b$=$0.048$, $\Omega _{\Lambda}$=$0.693$, $\sigma_8$=$0.8288$, and a Hubble constant of $H_0$=$100\,h\,\rm km\,s^{-1}\,\rm Mpc^{-1}$, where $h$=$0.6777$. The typical mass resolution contained in each Lagrangian volume for gas and DM particles is $m_{\rm gas}$=5$\times$$10^4$$\,\rm M_\odot$ and $m_{ \rm DM}$=3$\times$$10^5$$\,\rm M_\odot$, respectively. The co-moving softening length of collisionless particles was set to 500$\,h^{-1}\,\rm cpc$; the physical softening length was kept fixed to 250$\,h^{-1}\,\rm pc$ below $z$=$1$.
 
We set an imposed minimum wind velocity of $v_{\rm w,min}$=350$\,\rm km\,s^{-1}$ following \citep{Annalisa2018}, instead of the default value of $v_{\rm w,min}$=0$\,\rm km\,s^{-1}$ used in the original \texttt{AURIGA} physics model. For each of the ICs, we performed two sets of simulations; the first one included the whole physical formation model. We hereafter refer to this configuration as the ‘fiducial’ or ‘AGN’ run. In the second configuration, we ran the same galaxies, but without including BHs and related AGN feedback. We refer to this configuration as the ‘non-AGN’ run.  A detailed description of the simulation suite sample can be found in \citet{ArjonaGalvez2024}. 

Haloes and substructures are identified using a Friends-of-Friends (FoF) algorithm. Gravitationally bound substructures within each FoF group are subsequently identified using the \texttt{SUBFIND} algorithm \citep{Springel2001,Dolag2009}. For each FoF group, the most massive bound subhalo is defined as the central galaxy, while the remaining bound structures are classified as satellites. The centre of the main halo is taken to be the position of the particle with the minimum gravitational potential of the central subhalo. Virial quantities are computed within a spherical region of radius $R_{200}$, enclosing a mean density 200 times the critical density of the Universe at the corresponding redshift. The associated mass within this radius defines $M_{200}$. In this work, we focus exclusively on the galaxy H0 from \citet{ArjonaGalvez2024}, which exhibits the clearest outflow signatures, with a virial mass of M$_{200} = 2.7\times10^{11}$M$_\odot$, and a stellar mass of M$_{\star}=5.7\times10^{9}$M$_\odot$ at $z$=0, hosting the most massive BH in our sample, M$_{\rm BH}=1.3\times10^7$M$_\odot$.

\subsection{BH seeding, accretion and feedback}

BHs are seeded in haloes exceeding a friends-of-friends mass threshold of $5\times10^{10}\,{\rm M_\odot}\,h^{-1}$. The seed mass is $10^5\,{\rm M_{\odot}}\,h^{-1}$, consistent with expectations for massive Population III remnants or direct-collapse scenarios. Seeds are placed at the position of the densest gas cell within the halo.

Gas accretion onto BHs follows a Bondi–Hoyle–Lyttleton prescription \citep{Bondi1944,Bondi1952}, including modifications to account for unresolved cold gas and angular momentum effects. The AGN feedback model follows the implementation described in \citet{Weinberger2017}. In our simulations, feedback energy is injected thermally in the so-called quasar mode. The energy injection rate is given by

\begin{equation}
\Dot{E} = \epsilon_\mathrm{f}\,\epsilon_\mathrm{r}\,\Dot{M}_{\mathrm{BH}}\,c^2 ,
\end{equation}

\noindent where $\Dot{M}_{\mathrm{BH}}$ is the BH accretion rate, $\epsilon_\mathrm{r}$ the radiative efficiency, and $\epsilon_\mathrm{f}$ the coupling efficiency. The injected energy is distributed among neighbouring gas cells with a distance-weighted kernel. In addition, radiative feedback locally enhances the UV background, following \citet{Vogelsberger2013}. A detailed description of the adopted parameters can be found in \citet{Grand2017} and \citet{ArjonaGalvez2024}.

\begin{figure}[t!]
   \centering \includegraphics[width=0.9\columnwidth,trim={0 1cm 0 0.6cm}]{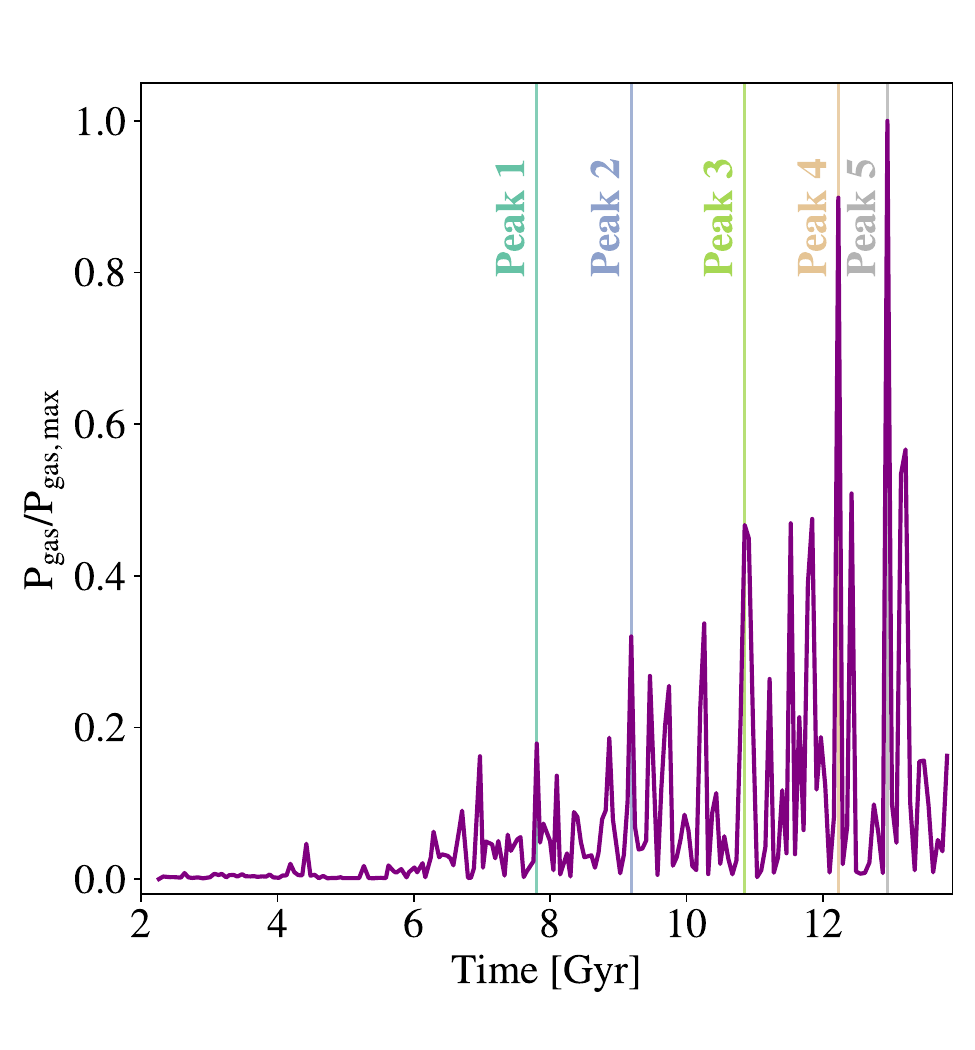}
    \caption{Normalised local gas pressure of the nearest gas cells around the BH as a function of cosmic time. Values are normalised by the maximum pressure over the full evolution. Vertical lines indicate selected outflow episodes associated with representative pressure peaks. These peaks correspond to hot, over-pressurised bubbles capable of driving AGN winds.}
    \label{fig:outflowcheck}
\end{figure}

\subsection{Outflow identification and characterisation}

To characterise the outflows produced in our simulations by AGN activity, we examine the energy deposited by the BH into the surrounding gas and its response over time. Specifically, we track the formation of quasar-heated bubbles in the vicinity of the BH by evaluating the local gas pressure, computed as

\begin{equation}
P = (\gamma - 1)\,\rho\,u ,
\end{equation}

\noindent where $\gamma$ is the adiabatic index of the gas, equal to 5/3, $\rho$ is the local comoving gas density, and $u$ is the local thermal energy per unit mass, both evaluated within the physical radius enclosing the 384$\pm$48 nearest gas cells around the BH. 

These over-pressurised regions appear as transient \textit{peaks} in the temporal evolution of gas pressure near the BH. Each peak corresponds to the rapid injection of energy by the AGN, resulting in hot bubbles capable of driving gas outflows. By inspecting the normalised pressure evolution, we can systematically identify individual outflow events as peaks that rise above the surrounding pressure level. In Fig.~\ref{fig:outflowcheck}, we show a subset of these peaks as examples, indicated with vertical lines; not all pressure peaks are marked, but the selected ones illustrate typical events capable of creating fast-hot gas bubbles that can be categorised as AGN-driven outflows (see Fig. \ref{fig:outflowsdiagram}).

We classified gas as outflowing if it meets two criteria: (i) it belongs to an over-pressurised region and (ii) it has a positive radial velocity relative to the centre of the galaxy. A stricter criterion is applied for escaping outflows, requiring

\begin{equation}
v_{\rm rad} > v_{\rm esc}(r),
\end{equation}

\noindent where $v_{\rm esc}(r)$ is the local escape velocity computed from the gravitational potential at radius $r$. The galaxy centre is defined as the location of the SMBH, providing a consistent reference for radial velocities.

To ensure that selected gas corresponds to pressure-driven outflow events, we require gas within 1~kpc of the BH to have a temperature $T > 10^{6}\,\rm K$ and a density $\rho > 10^{6}\,{\rm M_\odot\,kpc^{-3}}$ \citep[see][]{Irodotou2022}. These thresholds isolate hot, dense bubbles capable of surpassing the escaping velocity of the galaxy while minimising contamination from quiescent ISM gas. To follow the spatial and thermodynamic evolution of this outflowing material over time, we employ a Lagrarian particle tracking approach using the Monte Carlo tracer particles implemented in the \texttt{AREPO} code \citep{Genel2013}. These tracers are advected with the mass flow across the moving Voronoi mesh, allowing us to reconstruct the trajectories and physical history of the gas elements accelerated during individual AGN episodes. Altogether, this methodology allows us to systematically identify and characterise AGN-driven outflows, as illustrated in Fig. \ref{fig:outflowcheck}, which highlights the temporal correspondence between pressure peaks and outflow launches. A detailed analysis of the physical properties and the temporal evolution of these outflows is presented in Section \ref{sec:outflowproperties}.

\subsection{Synthetic emission modelling}

To connect the intrinsic physical properties of the simulated galaxies with observable diagnostics, we generate synthetic nebular emission using the \texttt{Synthesizer} framework \citep{Lovell2025,Roper2025}. \texttt{Synthesizer} is a modular package designed to translate physical quantities from cosmological simulations into directly observable spectra, photometry, and emission-line properties in a self-consistent manner. This code combines stellar population synthesis and photoionisation modelling to compute emission from composite systems including stars, gas, dust, and AGN. The nebular emission and line ratios are computed using pre-computed photoionisation grids generated with the \textsc{Cloudy} code \citep[e.g.,][]{Chatzikos2023}\footnote{We use the latest \textsc{Cloudy} c23.01-based grids embedded within \texttt{Synthesizer}.}, which provide the underlying ionisation balance and emissivity predictions for a wide range of physical conditions. In particular, the BPT diagnostics rely on these \textsc{Cloudy}-based grids to ensure a self-consistent mapping between gas physical properties and the resulting emission-line ratios. Finally, emission outputs are handled through dedicated spectral and line objects, allowing direct construction of diagnostic line ratios and comparison with observational classification schemes.

\subsubsection{Stellar emission}

Stellar particles are treated as individual ionising sources. Their spectral energy distributions (SEDs) are assigned according to their mass, age, and metallicity using a pre-computed stellar population synthesis grid. In this work, we assume a Chabrier initial mass function (IMF)\footnote{In particular, we adopt for the SPS the\\  \texttt{bc03-2016-Miles\_chabrier-0.1,100\_cloudy-c23.01-sps} grid.}, consistent with the IMF adopted in the \texttt{AURIGA} simulations, ensuring internal consistency between the stellar populations existing in our sample and those employed in the emission modelling. The resulting ionising photon production rate from stellar populations is coupled to the surrounding gas, whose density, temperature, and metallicity determine the emergent nebular emission.

\subsubsection{AGN emission}\label{sec:AGNmodel}

To incorporate the AGN component within \texttt{Synthesizer}, we extend the default simulation loader to include the BH population following the method of Matamoro Zatarain et al. (in prep.). In this approach, BH metallicities are estimated from the local environment, typically by averaging the metallicity of either the stellar or gas component within a characteristic radius of 1 kpc from the BH. This provides the necessary input for coupling BHs to the photoionisation modelling while maintaining consistency with the surrounding galactic ISM.

The accreting BH is modelled using the \texttt{UnifiedAGN} implementation available within \texttt{Synthesizer} (Vijayan et al., in prep.). In this framework, the AGN emission is generated from physically motivated models linking BH properties to the emitted spectrum. The bolometric luminosity is derived from the instantaneous BH growth rate according to

\begin{equation}
    L_{\rm BH,\mathrm{bol}} = \epsilon_{\rm r}\dot{M}_{\rm BH}c^{2},
\end{equation}

\noindent where $\dot{M}_{\rm BH}$ is the BH accretion rate and $\epsilon_{\rm r}$ is the radiative efficiency. For constant $\epsilon_{\rm r}$, this establishes a direct proportionality between accretion rate and bolometric luminosity.

In this work, we adopt the relativistic \texttt{qsosed} accretion disc model from \texttt{Synthesizer}, which explicitly links the emergent AGN spectrum to the BH mass and Eddington ratio. The disc emission illuminates both a narrow-line region (NLR) and a broad-line region (BLR). However, since our analysis focuses on classical narrow-line diagnostics, we restrict our modelling entirely to the NLR component, which is the primary contributor to the optical forbidden lines considered here.

The NLR is characterised by an ionisation parameter $U_{\rm NLR}$ given by

\begin{equation}
    U_{\rm NLR} = \frac{Q}{4\pi R_{\rm NLR}^2 n_H c}
    \label{eq:Unlr}
\end{equation}

\noindent were $n_{\rm H}$ and $R_{\rm NLR}$ refer to the hydrogen density and the radius of the NLR, $c$ refers to the speed of light, and $Q$ is the corresponding ionising photon produced per second by the central AGN. The intrinsic parameters needed to compute the NLR are modelled using pre-computed photoionisation grids \footnote{We make use of the NLR grid corresponding to \\ \texttt{relagn\_fixed\_rad\_efficiency\_0p1\_cloudy-c23.01-nlr.hdf5} grid.}. In practice, the ionisation parameter, $U_{\rm NLR}$, is determined by the incident ionising photon flux relative to the gas density. To compute the ionising photon budget from the AGN, we assume that a fraction $f_{\rm ion}=0.3$ of the bolometric luminosity contributes to photons with energies above 13.6 eV, giving an ionising luminosity of $L_{\rm ion}=f_{\rm ion}L_{\rm bol}$. This is converted to the photon rate $Q$ by dividing by the mean energy of the ionising photons. For BHs of mass $\sim10^{7}\,{\rm M_\odot}$, theoretical AGN SED models \citep{NetzerTrakhtenbrot2014} indicate that approximately 20–40\% of $L_{\rm bol}$ lies above the hydrogen ionisation threshold, consistent with observational constraints for AGN at Seyfert-like luminosities\footnote{While these sources have historically been classified as \textit{Seyfert} galaxies based on early optical observations, we will adopt the more general physical term \textit{AGN} throughout this work.} and with standard practice in NLR photoionisation modelling \citep{Feltre2016}. This ensures that variations in the simulated accretion rate are self-consistently reflected in the ionisation state of the NLR gas. We assume fixed $n_{\rm H}$ and $R_{\rm NLR}$ values of 10$^2$ cm$^{-3}$ and 300 pc, respectively, corresponding to low-mass galaxies with shallow gas potentials. These values are chosen such that the resulting NLR remains consistent with the ionisation-bounded assumption over the range of AGN luminosities considered in this work.

\subsubsection{Emission-line diagnostics}

For each snapshot analysed, we compute integrated emission-line luminosities for the transitions relevant to classical diagnostic diagrams: H$\alpha$, H$\beta$, [O\,III]$\lambda5007$, [N\,II]$\lambda6584$, [S\,II]$\lambda6717$, and [O\,I]$\lambda6300$. These are extracted using the \texttt{LineCollection} interface within \texttt{Synthesizer}, which allows direct calculation of line ratios and diagnostic classifications.

Because both stellar and AGN radiation fields are derived from the instantaneous properties of the simulation (stellar ages and metallicities, BH mass, and accretion rate), this approach naturally captures the evolution of excitation conditions across our simulation snapshots. The resulting BPT and VO87 diagrams therefore provide a physically grounded bridge between the simulated AGN duty cycle and its observable spectroscopic signatures.

\begin{figure*}[t!]   \includegraphics[width=\textwidth,trim=2cm 0 0 0,clip]{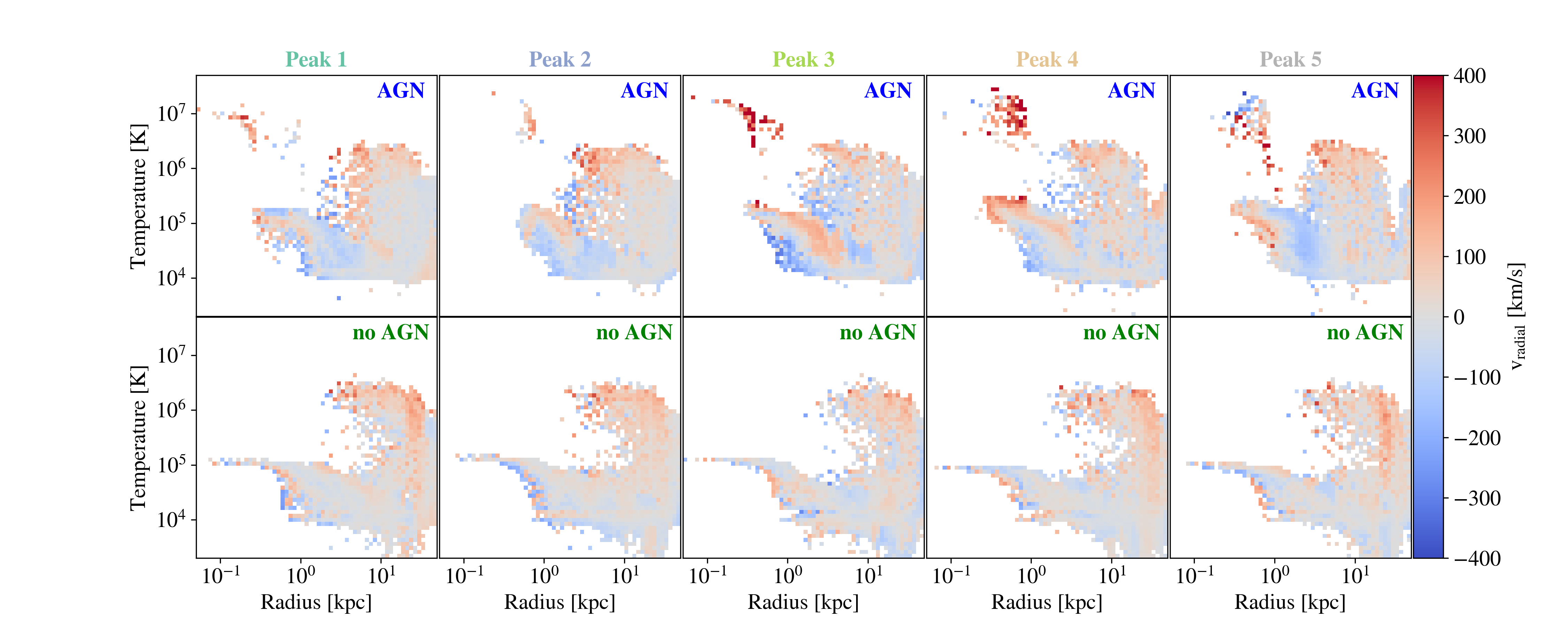}
    \caption{Temperature vs radius diagrams for gas associated with the identified pressure peaks in Fig. \ref{fig:outflowcheck}. Colour indicates radial velocity values. Top row: AGN simulation; bottom row: no-AGN counterpart, both at the same time. Over-pressurised bubbles with high outward velocities are only present in the AGN run.}
    \label{fig:outflowsdiagram}
\end{figure*}

\section{Results}

Understanding the physical properties of AGN-driven outflows is essential for quantifying the impact of BH feedback on galaxy evolution. Outflows regulate SF by removing or heating the ISM and redistributing metals across the galaxy halo, ultimately influencing the morphology and thermodynamics of the host galaxy \citep[e.g.][]{Fabian2012, KingNixon2015}. Despite extensive observational and theoretical efforts, the precise mechanisms controlling the velocity, temperature, density, and spatial extent of these outflows remain under debate, particularly for low-mass galaxies where stellar and AGN feedback may compete. In this work, we examine the properties of outflows in the central H0 galaxy, presented in \cite{ArjonaGalvez2024}, comparing the AGN and no-AGN samples, to isolate the effects of BH activity.

\subsection{Physical properties of the outflows}\label{sec:outflowproperties}

The temporal evolution of the local gas pressure near the BH (Fig.~\ref{fig:outflowcheck}) reveals a sequence of sharp, transient peaks. These peaks trace discrete episodes of energy injection associated with AGN activity and correspond to the formation of hot, over-pressurised bubbles. Fig. \ref{fig:outflowsdiagram} shows the gas properties for each of the five selected peaks, marked with a vertical line in Fig. \ref{fig:outflowcheck}. In the AGN simulation (top panels), the pressure enhancement corresponds to a compact region of gas heated to T$\sim$$10^6$–$10^8$K within the central hundreds of parsecs. This gas exhibits predominantly positive radial velocities, mostly exceeding the local escape velocity. In contrast, the no-AGN counterpart (bottom panels), analysed at the same time snapshot, lacks both the extreme temperatures and the high-velocity tail, demonstrating that stellar feedback alone does not produce comparable structures.

\begin{figure*}[t!]
\centering\includegraphics[width=0.9\textwidth,trim={0 3.5cm 0 4.4cm},clip]{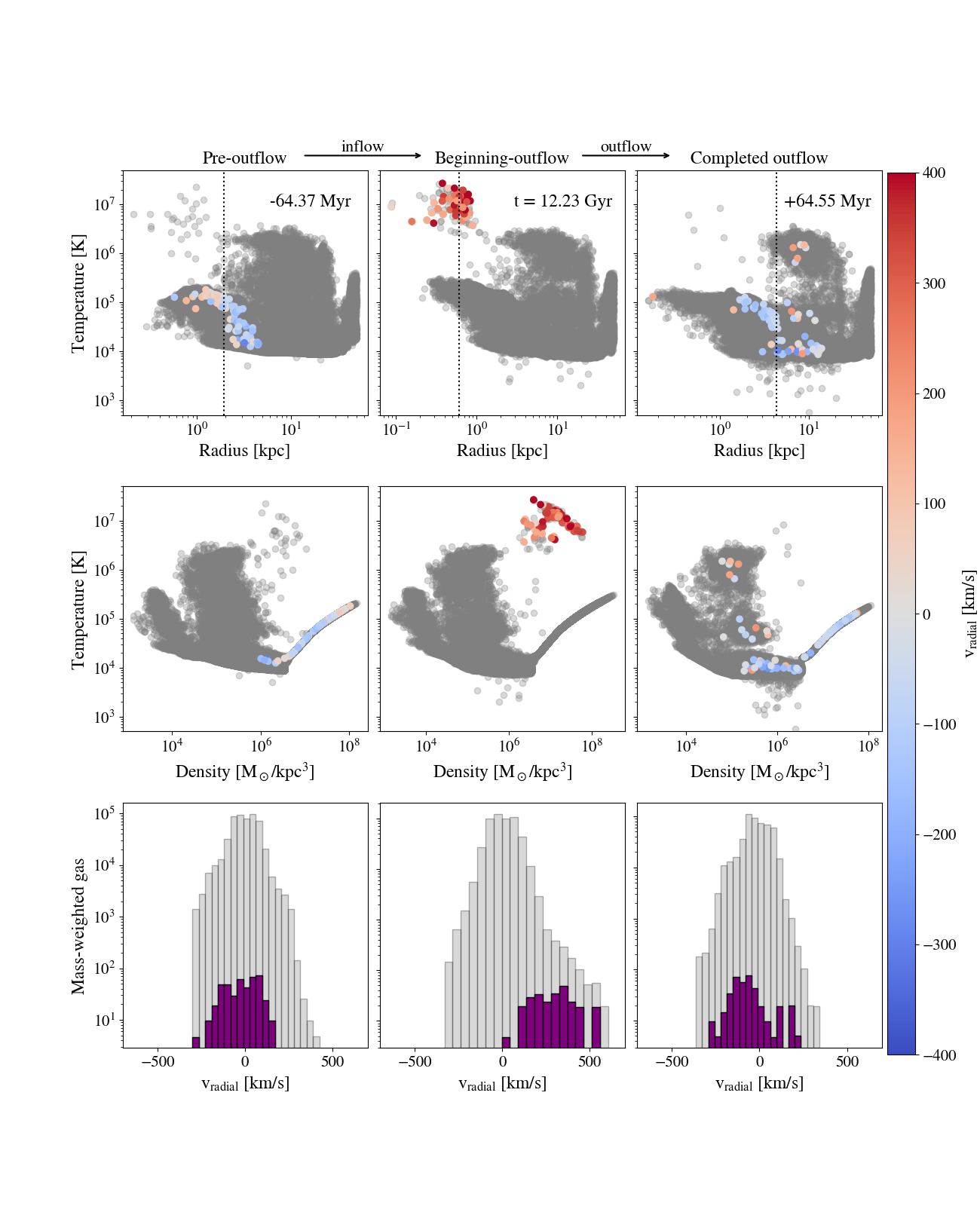}
    \caption{Outflowing gas at t$=12.23$ Gyr (\textit{peak 4} in Figs \ref{fig:outflowcheck} and \ref{fig:outflowsdiagram}) in our AGN-galaxy, colour-coded by the radial velocity of each tracked particle relative to the BH for the first and middle rows and in purple for the last row. The remaining gas is shown in grey. Top row: temperature vs radius for the gas at three times: one snapshot before the event, at the selected outflow event, and one snapshot after. In each panel, the vertical dotted line indicates the median radial position of the outflowing gas. Middle row: temperature vs density for the same gas, again, colour-coded by radial velocity. Bottom row: Radial velocity histograms of the outflowing and all gas, shown in purple and grey, respectively. The outflowing gas with T$>10^6$ K and $\rho$>$10^6$ M$\odot$/kpc$^3$ occupies the high-velocity tail (middle-column) and remains confined within $\sim$10~kpc thereafter (right-column), illustrating the recycling of hot gas.}
    \label{fig:track}
\end{figure*}

These AGN-driven bubbles occupy a regime of simultaneously high density and high temperature, with typical densities of 
$\rho$$\sim$$10^6$-$10^8$ M$_\odot$/kpc$^3$, comparable to star-forming gas, but at temperatures several orders of magnitude higher than the SF temperature (see Fig. \ref{fig:track}). Notably, this high-density, high-temperature gas confined to the central regions is consistent with recent observations by \cite{Aravindan2026}, who report that the coronal line region in dwarf AGN with outflows is similarly restricted to the inner 0.5 kpc. This creates a distinct branch in the phase space, absent in the no-AGN simulation. This thermodynamic configuration implies that AGN feedback in dwarfs does not simply evacuate low-density gas. Instead, it injects energy directly into dense central regions, producing over-pressurised structures that temporarily stabilise the gas against gravitational collapse. The elevated temperature increases the local Jeans mass and prolongs cooling times, thereby suppressing SF without requiring immediate mass removal. Interestingly, this preventive mode of feedback closely resembles the effect of the thermal quasar mode feedback described by \citep{Irodotou2022} for more massive, MW-like haloes within the \texttt{AURIGA} suite, suggesting a continuity in how central engines regulate the inner ISM across different galaxy regimes. 

From now on, we will focus our analysis on the event reaching $P/P_{max}\sim$0.9 at $t=12.23$ Gyr (\textit{peak 4)} in Fig. \ref{fig:outflowcheck} and \ref{fig:outflowsdiagram}, which represents a clean outflow episode dominated by outward motion. Note that we exclude the absolute maximum peak (\textit{peak 5}) as it is partially contaminated by artificial inflows linked to the BH repositioning scheme. However, we provide the corresponding maps and velocity analysis for \textit{peak 5} in Appendix \ref{app:peak5}, as it demonstrates that the overall physical trends remain consistent across different feedback episodes despite these numerical artefacts.

\begin{figure}[t!]
    \centering
    \includegraphics[scale=0.55]{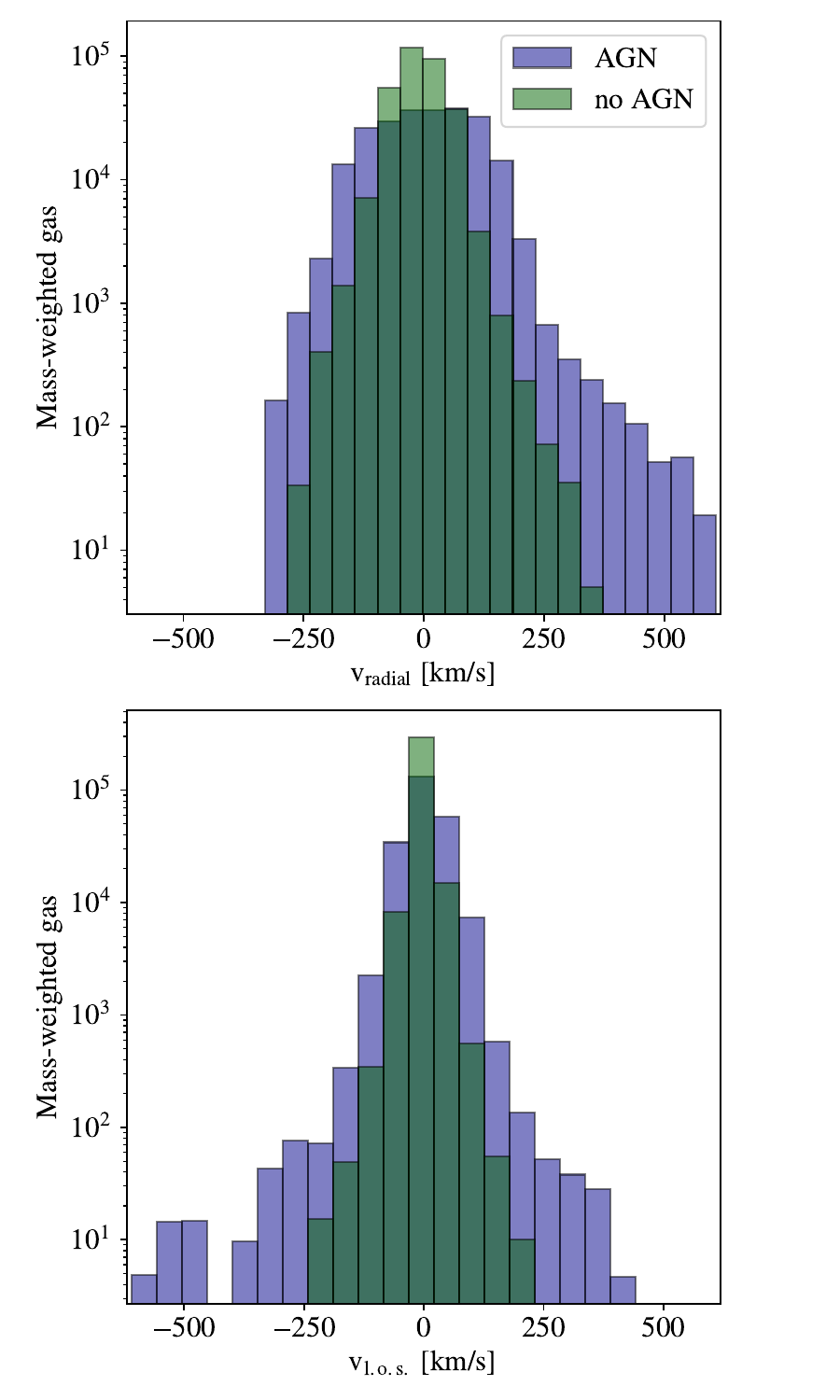}
    \caption{Radial velocity (top) and line-of-sight velocity (bottom) histograms of gas within 10\% of the virial radius for the AGN (blue) and no-AGN (green) simulations at $t=12.23$ Gyr, corresponding to the selected pressure \textit{peak 4} shown in Figs. \ref{fig:outflowcheck} and \ref{fig:outflowsdiagram}. The AGN outflow produces a high-velocity tail of v$_{\rm radial}$, extending to $\sim 600~\rm km\,s^{-1}$, whereas the no-AGN counterpart remains below $\sim 300~\rm km\,s^{-1}$. The l.o.s. distribution is broadened for the AGN case, reflecting also a velocity tail of $\sim$-500 km/s.}
   \label{fig:vel227}
\end{figure}

\begin{figure*}[t!]    \centering\includegraphics[width=0.9\textwidth,trim={2cm 1cm 2.5cm 0.5cm}]{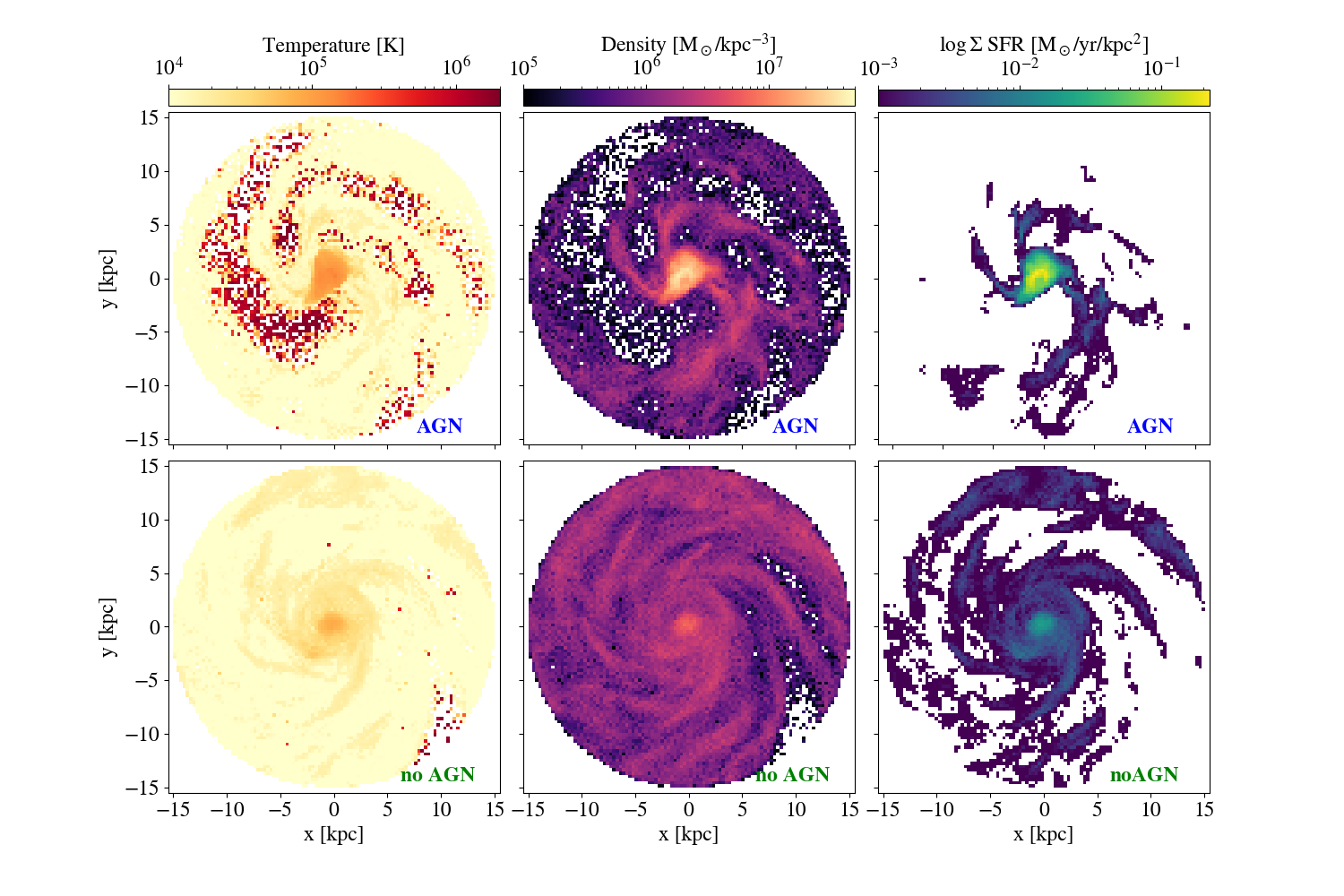}
    \caption{Face-on maps of H0 galaxy within 15~kpc of radius for the corresponding \textit{peak 4} at $t=12.23$ Gyr illustrated in Figs. \ref{fig:outflowcheck} and \ref{fig:outflowsdiagram}. From left to right: gas temperature, gas density, and surface SFR. Top row: AGN simulation; bottom row: no-AGN counterpart. The centre region of the AGN run shows hotter, denser gas and enhanced central SF compared to the no-AGN case, while the large-scale disk is highly fragmented.}
    \label{fig:map227}
\end{figure*}

\subsection{Temporal evolution and recycling of gas}
\label{sec:tracking}

To follow the evolution of the outflowing material, we perform a Lagrangian tracking analysis using the \texttt{AURIGA} tracer particles associated with the outflowing gas. Fig. \ref{fig:track} tracks the properties of the gas cells at $t=12.23$ Gyr associated with the pressure \textit{peak 4} shown in Figs. \ref{fig:outflowcheck} and \ref{fig:outflowsdiagram}. The top row displays the temperature–radius diagrams approximately 64 Myr before (left column) and 64 Myr after the selected episode (right column). We track those gas cells that exceed the escape velocity given by their radius and colour them by radial velocity at every snapshot. For context, all gas within 15 kpc in the x-y plane and $\mid$$z$$\mid < 2.5$kpc is represented in grey. Note that we are limited by the time resolution of our snapshots\footnote{This cadence should not be confused with the internal integration timestep of the simulation, which is about $\sim$$ 1.6\times10^{-2}\,\mathrm{Myr}$ at this epoch. The underlying hydrodynamical evolution, including cooling and feedback processes, is evolved on significantly smaller temporal scales than those sampled by the snapshots.}.  The snapshot spacing varies between $\sim 60\text{--}100\,\mathrm{Myr}$ across cosmic time and is precisely 64 Myr during this specific epoch, which restricts our tracking analysis to these discrete time between snapshots.

The middle row of Fig. \ref{fig:track} shows the corresponding temperature–density diagrams. Prior to the outflow (left column), the gas is concentrated within the central 5 kpcs, primarily occupying high-density, SF regions, and it begins to move inwards towards the inner 1 kpcs. During the selected event (middle column), the gas initiates the outflow, as it is rapidly heated to temperatures of $\sim$10$^7$ K and accelerated outward, populating the high-velocity tail of the radial velocity distribution (bottom panels). Notably, this outflowing gas does not follow the canonical SF equation of state that imposed an effective temperature \citep{SpringelHernquist2003}. Instead, it occupies a high-temperature, high-density regime, confirming that the bubble represents a thermodynamically distinct state produced by AGN energy injection.

The bottom panel presents the radial velocity distribution of the gas shown in the phase-space diagrams above. In grey, we represent the radial velocity histogram corresponding to the gas sample displayed in the temperature–radius and temperature–density panels, while the purple histogram highlights the subset of particles belonging to the selected outflowing bubble that satisfy $v_{rad}>v_{esc}$. This subset represents the fraction of the tracked gas that is accelerated beyond the local escape velocity during the outflow episode. The AGN episode significantly enhances the high-velocity end of the distribution. The subsequent evolution (right column) reveals that only a limited fraction of the affected material truly escapes the system. Although these particles initially exceed the local escape velocity, they quickly plough into the ISM and inner CGM, dissipating their kinetic energy and causing the gas to decelerate and return to the disc at most after 64 Myr. This demonstrates that the dominant mode of AGN feedback in this stellar-mass regime is not large-scale gas expulsion, but rather internal redistribution and rapid gas recycling back into the galactic disc, a fountain-like behaviour conceptually similar to the feedback cycles described for MW-like galaxies within the \texttt{AURIGA} suite \citep{Grand2019}.

While stellar feedback commonly drives global, long-timescale fountain flows in dwarfs \citep{Brook2012}, in the \texttt{AURIGA} model, AGN activity introduces a distinct, highly energetic nuclear channel. It generates localized thermal bubbles of hot, dense gas that expand, cool, and efficiently remix into the central ISM within less than $\sim$$64\,\mathrm{Myr}$. This rapid cycle significantly impacts the chemical structure of the nucleus; as shown in Appendix \ref{app:ZAGNnoAGN}, the AGN configuration produces a broader overall metallicity distribution and a lower characteristic metallicity within the central $0.1\,R_{200}$ due to enhanced mixing between the nuclear gas and the surrounding ISM. While global stellar feedback regulates the global ISM over longer timescales, it may not always reach the extreme thermodynamic states required to lift and rapidly redistribute the densest core reservoirs. Consequently, alongside stellar winds, episodic AGN-driven feedback plays a key role in managing the nuclear gas supply through rapid, localized recycling rather than permanent gas expulsion.

\subsection{Kinematics of the outflows}
\label{sec:kinematics}

The global kinematic impact of AGN feedback is illustrated in Fig. \ref{fig:vel227}, which compares the gas velocity distributions for the AGN (blue) and no-AGN (green) simulations during the pressure event \textit{peak 4} shown in Figs. \ref{fig:outflowcheck} and \ref{fig:outflowsdiagram}. To capture the dynamics of the ISM and the inner circumgalactic medium (CGM), we consider all gas within 10\% of the virial radius.

In the radial velocity distribution (top panel), the AGN-hosting galaxy exhibits a prominent high-velocity tail extending up to $\sim$600 km s$^{-1}$ at outflow time. This feature is entirely absent in the no-AGN run, where gas velocities rarely exceed 200-300kms$^{-1}$. This contrast aligns with the findings of \citet{Koudmani2019}, who found that while supernova feedback is efficient at regulating SF through mass-loading, it cannot produce the fast and energetic outflows that emerge when AGN feedback is coupled to the ISM. Our results confirm that AGN episodes are the primary driver of the high-velocity gas phase in dwarfs, reaching speeds that can exceed the local escape velocity even in relatively shallow potential wells \citep[see also][]{Dubois2015,Koudmani2021}.

The line-of-sight (LOS) velocity distribution (bottom panel) provides the critical link to spectroscopic observations. While the no-AGN simulation shows a narrow, symmetric distribution centred at zero, AGN activity results in a substantially broader and asymmetric profile, with wings reaching |v$_{\rm l.o.s.}$| > 400 km s$^{-1}$. These high-velocity LOS components represent the kinematic signatures identified observationally as broad emission-line wings or blueshifted absorption features, matching detections of ionised outflows in local dwarf galaxies \citep[e.g.,][]{ManzanoKing2019, Liu2020, Aravindan23}. Notably, this shift from a symmetric SF profile to an asymmetric one with enhanced blue wings closely aligns the observational trends reported by \citet{Aravindan23}, who found that outflows in active dwarfs are systematically faster and more blueshifted than those driven purely by SF.

Despite these high velocities, our analysis reinforces the idea that the outflow is not a large-scale "blowout" event. Although most of the gas satisfies v$_{\rm rad}$ > v$_{\rm esc}$, the majority of the material remains gravitationally bound. As discussed in Section \ref{sec:tracking}, the rapid deceleration and return to the disc before $\sim$60 Myr suggest that the AGN generates compact, energetic winds that redistribute gas internally. This mode of feedback is less destructive than models with higher coupling efficiencies, explaining why these galaxies maintain their central gas reservoirs despite recurring BH activity \citep[consistent with the low column densities of such outflows in the CGM reported by][]{Ortame2026}.

\subsection{Spatial structure and impact on the central regions}\label{sec:maps}

The spatial morphology of the outflow and its immediate impact on the ISM across the disk are illustrated by the face-on maps in Fig. \ref{fig:map227} at the outflow time ($t=12.23\,\rm Gyr$). In the AGN simulation (top panels), the presence of high-temperature gas ($>10^6$ K) extends out to several kiloparsecs, creating a highly structured and patchy ISM compared to the smooth gas distribution of the no-AGN run (bottom panels).

Within the inner $\sim$2 kpcs, the AGN run exhibits a pronounced central concentration of dense gas (middle column) that directly correlates with enhanced SFR (right column). This configuration reflects the cumulative impact of the AGN feedback over cosmic time. While continuous energy injection maintains a globally hotter gas reservoir compared to the no-AGN counterpart, discrete accretion episodes trigger central over-pressurised bubbles (see Fig. \ref{fig:outflowsdiagram}). As these bubbles expand, they violently compress pockets of the remaining inner gas. In the context of the sub-grid model of \cite{SpringelHernquist2003}, this episodic compression pushes the gas above the critical density threshold for SF, lowering the local Jeans mass and temporarily enhancing the central SFR through an in situ central starburst. These results provide a physical realisation of the dual-mode feedback framework, where AGN-driven outflows can simultaneously suppress global SF while locally enhancing it through gas compression and fragmentation \citep[e.g.,][]{SilkNusser2010,ZubovasBourne2017}. These findings are in agreement with recent cosmological simulations of more massive systems ($M_* \sim 10^{10}\,\rm M_\odot$) showing increased SF regions in quasars using \texttt{FIRE-2} simulations \citep{MercedesFeliz2023}. Furthermore, this process offers a theoretical counterpart to observational evidence of triggered SF in dwarf galaxies, such as the jet-induced activity seen in Henize~2-10 \citep{Schutte&Reines2022}.

On larger scales (out to $15\,\rm kpc$), both simulations preserve a well-defined galactic disk, but they differ fundamentally in the spatial uniformity of their properties. The no-AGN simulation maintains a smooth, continuous disk where gas density, temperature, and star formation smoothly trace the underlying spiral structure with gradual transitions. In contrast, the AGN-hosting galaxy displays much more abrupt changes. The cumulative injection of thermal energy from the central SMBH heats the gas non-uniformly. Consequently, the transitions between the dense, SF gas and the surrounding hotter, diffuse medium become significantly more sudden and irregular, resulting in a highly contrasted multi-phase distribution.

Ultimately, Fig. \ref{fig:map227} shows at $t=12.23\,\rm Gyr$ the non-trivial outcome of AGN feedback on this dwarf galaxy: enhanced star formation in the central region combined with reduced SF in the outskirts. Crucially, these structural differences are maintained down to $z=0$.

\begin{figure*}[t!]
\centering\includegraphics[width=\textwidth,trim={0 1cm 0 1cm}]{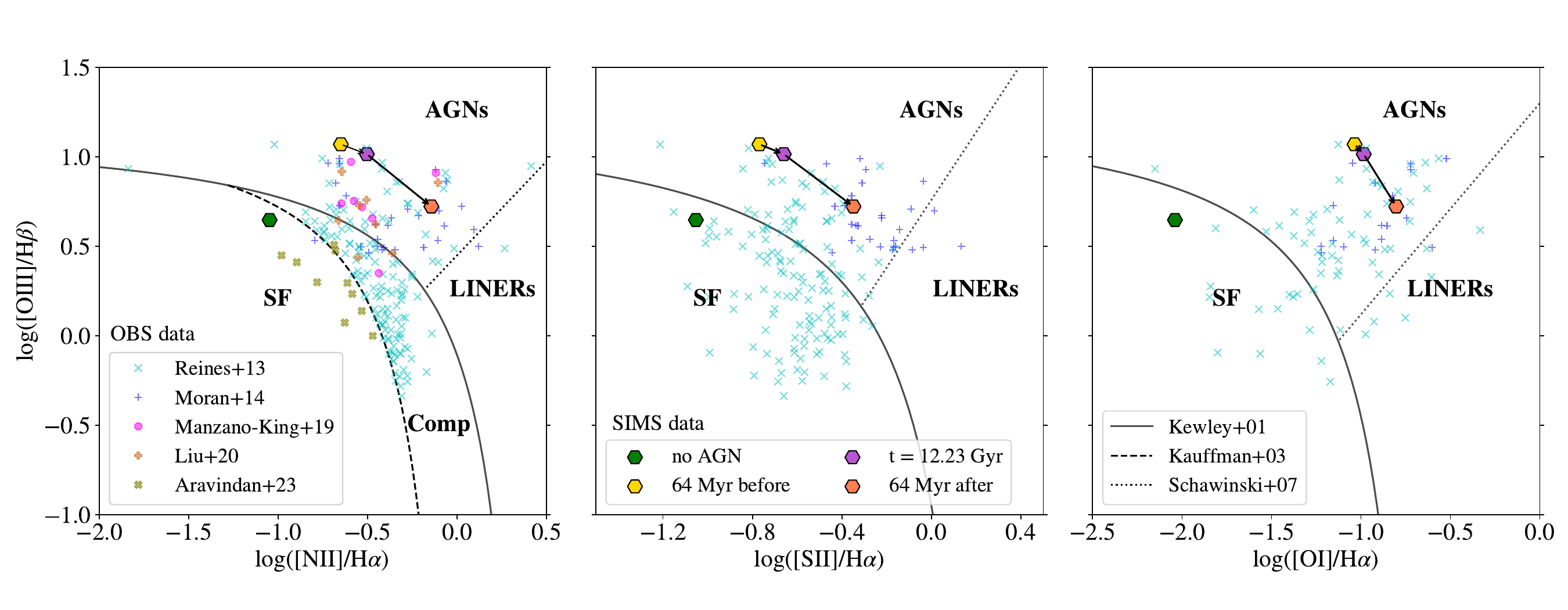}
    \caption{BPT narrow-line diagnostic diagrams for our simulated galaxy before, during and after the selected outflow event at $t=12.23$ Gyr (shown in Figs. \ref{fig:outflowcheck} and \ref{fig:outflowsdiagram} as \textit{peak 4}). Each of them shown with yellow, purple and orange hexagonal markers, respectively. In green, we also show the noAGN counterpart at the same outflow time ($t=12.23$ Gyr). The \cite{Kewley2001} starburst classification line (shown in solid), the \cite{Kauffmann2003} SF line (dashed line), and the AGN–LINER line from \cite{Kewley2006} (dotted line) are used to separate galaxies into  SFs, AGNs, LINERs, and composite SF–AGN types. In the BPT diagram (left panel), we show in orange crosses observations of AGN dwarfs from \cite{Liu2020} and with magenta circles observations from \cite{ManzanoKing2019}, indicating dwarf AGN with outflows. For comparison, we also indicate SF dwarfs from \cite{Aravindan23}. In addition, observations of dwarf AGNs from \cite{Moran2014} and \cite{Reines2013} are also shown for the BPT and VO87 (middle and left panels) diagrams using blue '+' and cyan 'x' markers. In each of the three diagnostic diagrams the AGN galaxy is well separated from its noAGN counterpart, lying in the expected observational region of the plot.}
    \label{fig:BPT}
\end{figure*}

\subsection{Synthetic observational signatures}
\label{sec:obscomparison}

To directly connect the intrinsic physical properties of the simulated AGN galaxy with observable diagnostics, we compute the synthetic narrow-line emission using the \texttt{UnifiedAGN} implementation available within \texttt{Synthesizer} (Vijayan et al., in prep.). In this section, we analyse the predicted optical emission-line ratios during a specific outflow event and across the entire evolutionary history of the dwarf galaxy since the formation of its central BH.

\subsubsection{Individual outflow episodes}

Fig. \ref{fig:BPT} presents the standard BPT diagrams for three distinct snapshots tracking the primary outflow event analysed in the previous sections: 64 Myr before the outflow (yellow markers), during the outflow event (purple markers), and 64 Myr after (orange markers). For comparison, the no-AGN counterpart at the epoch of the outflow of its counterpart AGN sim is shown in green.

The simulated AGN-hosting galaxy exhibits a clear, time-dependent trajectory through the diagnostic planes. Prior to the outflow, the BH accretion rate is relatively high, placing the galaxy firmly in the AGN region, displaying the highest [O\,III]/H$\beta$ ratios. As the outflow expands, the emission shifts toward higher [N\,II]/H$\alpha$, [S II]/H$\alpha$, and [O\,I]/H$\alpha$ ratios while the [O\,III]/H$\beta$ ratio declines, though it remains within the AGN-dominated regime. Finally, 64 Myr after the episode, the galaxy moves further toward lower excitation values, approaching the LINER region.

\begin{figure*}[t!]
\centering
\includegraphics[width=0.9\textwidth,trim={0 1cm 0 0}]{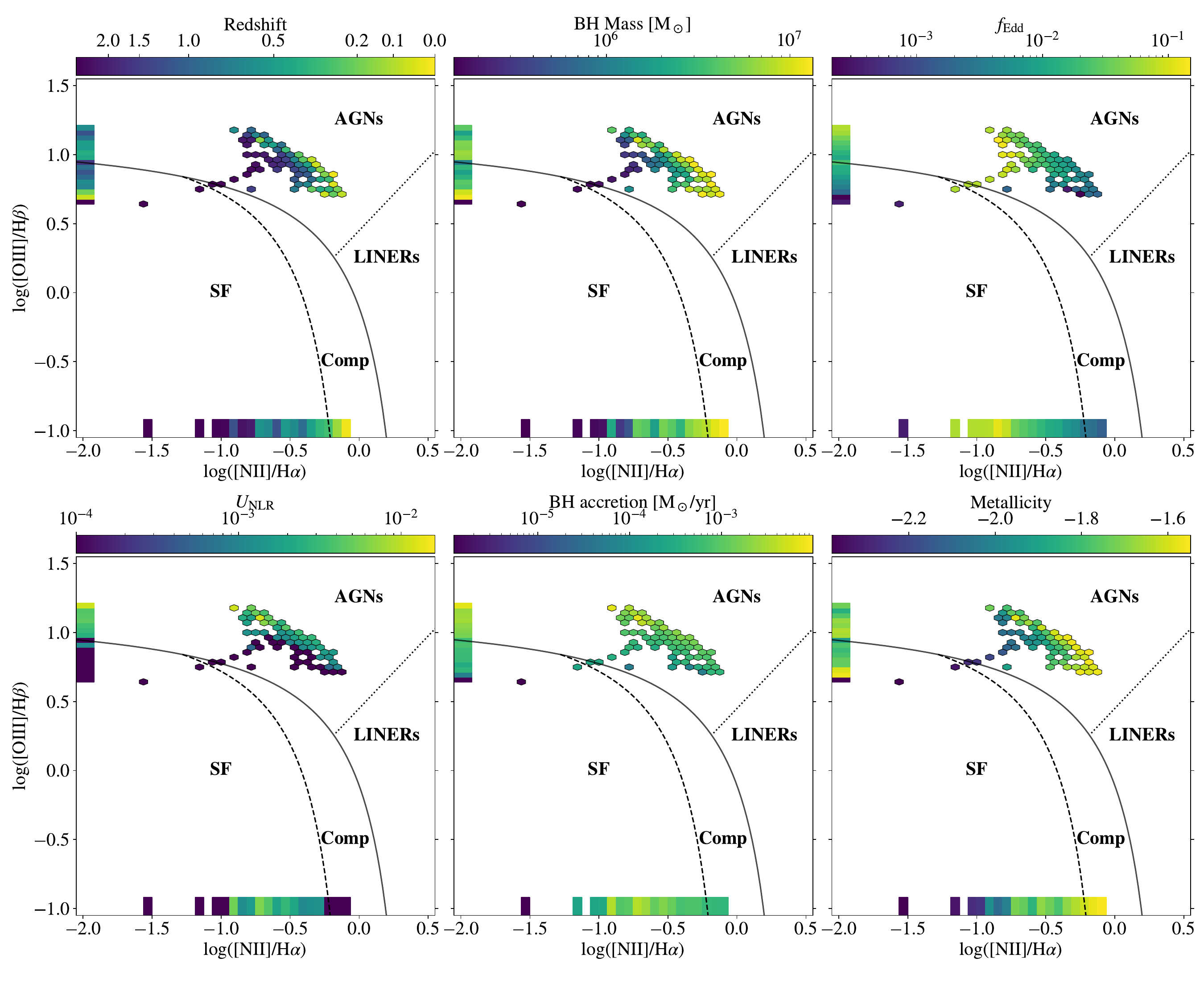}
    \caption{Evolution of our simulated AGN-galaxy across the  BPT diagram. The panels display 2D histogram of the narrow-line emission ratios $\rm \log([OIII]/H\beta)$ versus $\rm \log([NII]/H\alpha)$. Each hexagonal bin is colour-coded by the mean value of a specific physical quantity, indicated by the colorbars, tracing the evolution of the AGN. From left to right and top to bottom, the BPT diagram is colored by redshift, BH mass, Eddington ratio, NLR ionisation parameter, BH accretion rate, and metallicity. 
    The solid line represents the \cite{Kewley2001} starburst classification limit, the dashed line is the \cite{Kauffmann2003} empirical SF boundary, and the dotted line distinguishes between AGNs and LINERs \citep{Kewley2006}. These lines partition the diagnostic space into SF, Composite (Comp), AGN and LINER regions. Marginal coloured bands along the axis illustrate how the corresponding intrinsic parameter is distributed along the individual emission line ratios. The AGN galaxy departs from the SF space, as the BH grows and undergoes accretion episodes, it progressively moves toward the composite region, reaching the AGN location at $z$$\sim$2.6.}
    \label{fig:BPTevol}
\end{figure*}

This spectroscopic evolution is driven by the rapid decline in the central AGN radiation field as the feedback episode heats and displaces the inner gas, temporarily blocking the BH accretion. Since the ionisation parameter scales with the ionising photon flux and therefore with the AGN bolometric luminosity, which in turn depends on the BH accretion rate, the progressive reduction in the [O III]/H$\beta$ ratio directly reflects this weakening AGN radiation field. Specifically, the NLR ionisation parameter decreases monotonically from $U_{\rm NLR}=0.0209$ before the event, to $U_{\rm NLR}=0.0119$ during the outflow, and drops significantly to $U_{\rm NLR}=0.00188$ in the post-outflow phase. This tightly tracks the corresponding decline in the Eddington ratio from $f_{\rm Edd}=0.008$ to $0.005$ and finally down to $0.001$.

Comparing these predictions with observations, the simulated emission-line ratios overlap with the positions occupied by observed dwarf AGN from \cite{Liu2020} and specifically with ionised outflows from \cite{ManzanoKing2019}. These systems predominantly lie in the AGN region, similar to the points in our models, indicating that the excitation conditions produced by AGN activity in dwarfs is consistent with what is observed for outflowing systems in dwarf galaxies. The broader samples of dwarf AGN from \cite{Moran2014} and \cite{Reines2013} span both AGN and LINER classifications, overlapping with the range covered by the simulated galaxy across its evolutionary sequence.

\subsubsection{Evolution across the AGN life cycle}

While the three time frames in Fig. \ref{fig:BPT} illustrate the immediate impact of a single AGN-driven outflow episode, they represent only a tiny fraction of the highly variable AGN duty cycle. To place these events in a broader evolutionary context, Fig. \ref{fig:BPTevol} displays the full trajectory of our AGN-chosen galaxy in the [N\,II]-BPT plane from the time of BH seeding down to $z$=0. The tracks are coloured by key intrinsic parameters of the emission lines, such as BH mass, Eddington ratio, NLR ionisation parameter, and NLR metallicity within 1 kpc of the BH.

The different panels of Fig.~\ref{fig:BPTevol} show that the position of the galaxy in the BPT diagram evolves gradually over cosmic time at different radial scales. The redshift colouring highlights that these shifts correspond to different stages of the galaxy’s evolution. When the BH is initially seeded, the galaxy lies within the SF region of the diagram. As the BH grows and undergoes accretion episodes, the system progressively moves toward the composite region and eventually reaches the AGN location. Although the exact trajectory depends on the instantaneous gas conditions and the accretion state of the BH, the figure illustrates that the galaxy can transition between different BPT classifications over time. Notably, once the system transitions into the AGN regime, it remains there for the rest of its evolution. This persistent AGN classification is largely a consequence of the sub-grid physics implemented in the simulation. When examining the panels coloured by BH mass, Eddington ratio, and BH accretion rate, a general trend also emerges in which larger BH masses are typically associated with lower Eddington ratios and lower instantaneous accretion rates. As the BH grows, its feedback may become increasingly effective at removing or heating the surrounding gas, thereby reducing the available fuel for further accretion.

It is important to note that some parameters occasionally fall outside the limits of the photoionisation grid used to compute the emission-line ratios, and are therefore restricted by the boundaries of the adopted models. This is particularly evident for the NLR ionisation parameter, $U_{\rm NLR}$, defined previously by Eq. \ref{eq:Unlr}. Several snapshots reach the minimum value allowed by the grid (<$10^{-3}$), implying that these points should be interpreted as upper limits on the true value. In addition, the adopted NLR models assume a quasar-like ionising spectral energy distribution as implemented in \texttt{RELAGN}. While the Eddington ratios in the simulation fall within the nominal range covered by the grid, \texttt{RELAGN} does not explicitly account for potential changes in the ionising continuum shape at very low accretion states ($f_{\rm Edd} \lesssim 10^{-2}$). In this regime, the NLR properties would likely require a different modelling approach.

Another modelling choice concerns the metallicity adopted for the NLR. In Fig.~\ref{fig:BPTevol}, we use the mean stellar metallicity within 1 kpc of the BH, which reflects the inherited metallicity of the SF gas from which the stars formed and provides a relatively smooth tracer of the chemical enrichment of the nuclear region. The gas metallicity in the ISM can instead vary more strongly due to inflows, feedback, and mixing processes. In Appendix \ref{app:BPTevols}, we present the evolution of the mean stellar and gas metallicities over time (Fig. \ref{fig:metals}), together with alternative versions of Fig.~\ref{fig:BPTevol} where the NLR is modelled using the gas metallicity (Fig. \ref{fig:BPTgasZ}). Although the resulting BPT evolution appears more bursty when using the gas metallicity, the overall trends remain similar. Ideally, the metallicity adopted for the NLR would correspond to that of the gas actually accreted by the BH. However, this quantity is difficult to determine directly in the simulation. 

Despite these limitations and modelling assumptions, the simulated galaxy occupies regions of the BPT diagram that are comparable to those populated by observed systems. Therefore, even with the caveats discussed above, the model is able to place the simulated galaxy within the same general parameter space as observational samples, suggesting that the overall evolutionary behaviour captured in the simulation remains physically plausible.

\section{Conclusions}\label{sec:conclusions}

The presence of actively accreting BHs in dwarf galaxies has raised the possibility that AGN feedback may influence the evolution of galaxies well below the classical mass scale at which such processes are typically considered. In our previous work \citep{ArjonaGalvez2024}, we analysed a sample of dwarf galaxies from the \texttt{AURIGA} zoom-in simulations and showed that AGN activity can significantly affect their gas content and SF histories. In this paper, we focus on the galaxy hosting the most massive BH in that sample (H0, a dwarf galaxy with a $z$=0 stellar mass of $\sim$10$^{9.7}\,\rm M_\odot$ hosting a central BH of $\sim$10$^7\,\rm M_\odot$) in order to investigate in detail the physical properties, evolution, and observational signatures of AGN-driven outflows. By combining particle tracking with synthetic emission-line modelling, we connect the dynamical evolution of the outflowing gas with its potential observable signatures.

Our main results can be summarised as follows:

\begin{itemize}

\item AGN activity in the dwarf regime is characterised by sharp pressure peaks (Fig.~\ref{fig:track}) that drive the formation of hot, T$>10^{6}$ K, and dense, $\rho>10^{6}$ M$_\odot$/kpc$^3$, gas bubbles in the central kiloparsec of the galaxy (Figs.~\ref{fig:outflowsdiagram} and \ref{fig:track}). These over-pressurised structures represent a thermodynamically distinct phase of the ISM that is absent in simulations without an AGN, where stellar feedback alone is unable to reach such extreme temperatures and pressures in the dense central gas;

\item The remarkably rapid recycling timescale found in Fig. \ref{fig:outflowsdiagram} (less than 128 Myr, limited by the temporal resolution of our snapshot) contrasts sharply with SN-blastwave models implementing a temporary cooling shut-off, which typically predict prolonged fountain cycles spanning $\sim$1 Gyr to several Gyr for a fraction of the recycled material \citep{Brook2012}. 
Such short dynamical lifetime of individual outflows has to be taken into account when looking at the incidence of outflows reported in AGN dwarf-galaxy surveys (i.e., \cite{ManzanoKing2019} report 1/3 of their AGN showing outflow signatures);

\item Lagrangian particle tracking reveals that these bubbles can accelerate gas up to $\sim$$600\,\rm km\,s^{-1}$ (Fig.~\ref{fig:track} and \ref{fig:vel227}), significantly exceeding the velocities present in the corresponding simulation without an AGN. Although a fraction of this gas temporarily exceeds the local escape velocity, the vast majority of the material remains gravitationally bound within 10 kpc.  
The outflowing gas is heated and displaced before rapidly decelerating due to its interaction with the surrounding medium, falling again towards the center of the galaxy (Fig. \ref{fig:track}). This behaviour suggests that the dominant effect of the outflow is to heat, perturb, and regulate the central gas reservoir through a rapid recycling cycle, rather than to produce gas evacuation;

\item The spatial distribution of temperature, density, and SFR shown in Fig.~\ref{fig:map227} reveals that the AGN-driven outflows develop a complex structure as they propagate through the ISM. In particular, the hot gas preferentially expands along the lowest-density regions of the surrounding medium, creating anisotropic channels through which the outflow propagates more efficiently. As a result, the hot phase becomes distributed along these low-density pathways, the mechanical pressure of this anisotropic expansion compresses the surrounding colder gas, which triggers localised positive feedback, maintaining a dense, actively SF nuclear core that is absent in the no-AGN scenario;

\item The forward-modelled narrow‑line emission, computed with \texttt{Synthesizer} using the assumptions specified in section \ref{sec:AGNmodel}, demonstrates that AGN‑driven outflows in a dwarf galaxy imprint a clear, time‑dependent signature on classical diagnostic diagrams. In the BPT plane (Fig.~\ref{fig:BPT}), the galaxy moves inside the AGN region (64Myr before the outflow, yellow hexagon) towards higher [N\,II]/H$\alpha$, [S\,II]/H$\alpha$ and [O\,I]/H$\alpha$ ratios during the outflow (purple hexagon), approaching the LINER region after the episode (orange hexagon). This evolution mirrors the decline of the ionisation parameter and the Eddington ratio as the BH accretion weakens from 0.005 to 0.003 and finally down to 0.0005 M$_\odot$/yr. The synthetic points are consistent with the observed dwarf AGN from \cite{Liu2020} and dwarf-AGN outflows of \cite{ManzanoKing2019} and the broader dwarf samples of \cite{Moran2014} and \cite{Reines2013}. Conversely, the simulated noAGN counterpart remains in the SF region of the diagram (see green hexagon in Fig. \ref{fig:BPT}); 

\item The analysis of the simulated galaxy's evolution across the BPT diagram (Fig. \ref{fig:BPTevol}) reveals a clear trajectory over cosmic time. The system migrates from the SF region at high redshift through the composite region and eventually into the AGN regime. This progression demonstrates that as the BH grows more massive, its feedback becomes increasingly effective at altering the central environment, which generally leads to a decrease in both the accretion rate and the Eddington ratio. This indicates a strong self-regulation mechanism where the central engine limits its own growth by repeatedly disrupting its fuel supply.
\end{itemize}

Taken together, these results indicate that AGN-driven outflows in low-mass galaxies do not primarily quench SF through efficient gas removal. Instead, they act as a mechanism that periodically heats, accelerates, and redistributes gas within the central regions of the galaxy. This behaviour is consistent with the global trends identified in \citep{ArjonaGalvez2024}, where AGN feedback was shown to influence the gas content and SF activity of dwarf galaxies without necessarily expelling large amounts of material from their halos. Moreover, in this particular example, a global reduction of SF is accompanied by a local enhancement of SF in the central few kpc of the galaxy. We note that the results presented here correspond to a single system, hosting the most massive BH in our simulated dwarf galaxy sample, and that these findings are dependent on the specific AGN feedback scheme implemented in the code. Extending this analysis to different state-of-the-art AGN feedback models will be essential to establish the conditions under which AGN feedback becomes dynamically dominant in low-mass galaxies. Furthermore, extending the \texttt{Synthesizer} grids to fully cover the wide variability of the AGN regime remains limited by the assumed AGN spectral energy distribution in the \texttt{RELAGN} model, which may become physically uncertain at very low Eddington ratios where the model assumptions underlying the inner accretion flow structure break down. This limitation should be taken into account when interpreting the lowest-accretion states in the BPT diagram evolution of low-mass systems.

\begin{acknowledgements}
The authors sincerely thank Dr. Chris Lovell and Dr. Ana Contreras Santos for their valuable insights. E. Arjona-Gálvez acknowledges support from the Agencia Espacial de Investigación del Ministerio de Ciencia e Innovación (\textsc{AEI-MICIN}) and the European Social Fund (\textsc{ESF+}) through a FPI grant PRE2020-096361. ADC acknowledges financial support from the Spanish Ministry of Science
and Innovation (MICINN), through the \textit{Consolidación Investigadora} program, grant number CNS2023-144669, project “TINY”, and the \textit{Proyectos de Generación de Conocimiento} 2024 call, grant number PID2024-160009NA-I00, project
“INGENIO”. RJJG is supported by an STFC Ernest Rutherford Fellowship (ST/W003643/1). LVS acknowledges support from NSF-CAREER-1945310 and NSF-AST-2408339 grants. TMZ acknowledge support by the UKRI AIMLAC CDT, funded by grant EP/S023992/1. APV acknowledge support from the Sussex Astronomy Centre STFC Consolidated Grant (ST/X001040/1).

This research made use of the \texttt{LaPalma} HPC cluster, under project \texttt{can43}, PI A. Di Cintio, as well as the High-Performance Computers located at the Instituto de Astrofísica
de Canarias. The authors thankfully acknowledge the technical
expertise and assistance provided by the IAC (Servicios Informáticos Específicos, SIE) and the Spanish Supercomputing Network (Red Española de Supercomputación, RES).
\end{acknowledgements}

%
%

\bibliographystyle{aa}
\bibliography{ArjonaGalvez}

\begin{appendix}

\noindent\begin{minipage}{\textwidth}
 \section{Gas-phase metallicity distributions}
\label{app:ZAGNnoAGN}

Figure~\ref{fig:gasmetals} compares the mass-weighted gas-phase metallicity distributions of the AGN and non-AGN simulations. We show the distributions for all gas within $R_{200}$, as well as for the inner ($r<0.1R_{200}$) and outer ($r>0.1R_{200}$) regions separately.\\

The AGN run exhibits a broader metallicity distribution than the non-AGN counterpart, particularly in the low-metallicity tail. In the central region ($r<0.1R_{200}$), the gas is systematically shifted toward lower metallicities relative to the non-AGN case, while the outer region displays a wider spread in metallicity. These differences are consistent with enhanced gas circulation and mixing driven by AGN activity, which redistributes gas between the nuclear reservoir and larger galactic scales. Although the metallicity distributions alone do not uniquely determine the underlying transport mechanism, they provide independent support for the recycling scenario discussed in Section \ref{sec:tracking}.\\

\includegraphics[width=0.9\textwidth]{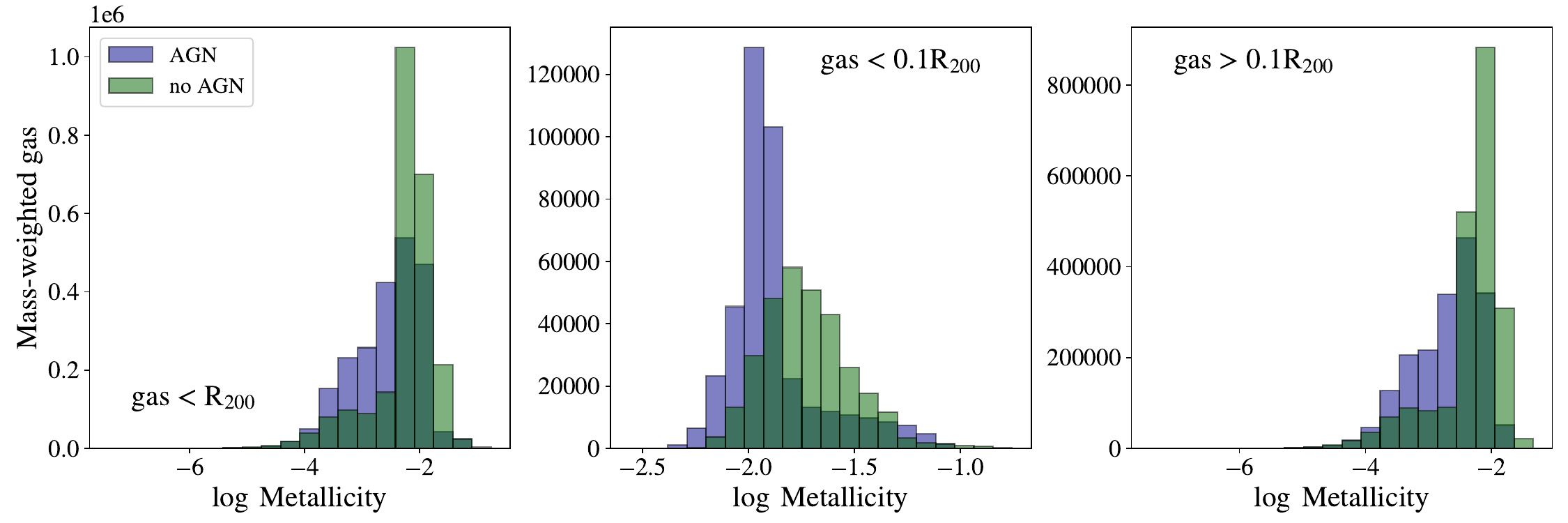}
\centering
\captionof{figure}{Gas metalicitty histograms for the AGN (blue) and no AGN (green) galaxy corresponding to \textit{peak 4} Figs. \ref{fig:outflowcheck} and \ref{fig:outflowsdiagram}. From left to right, the metallicity has been measure for the whole galaxy within R$_{200}$, within 0.1R$_{200}$ and outside R$_{200}$, respectively.}
\label{fig:gasmetals}

\section{Outflow properties of the maximum  pressure peak}
\label{app:peak5}

\includegraphics[width=0.9\textwidth]{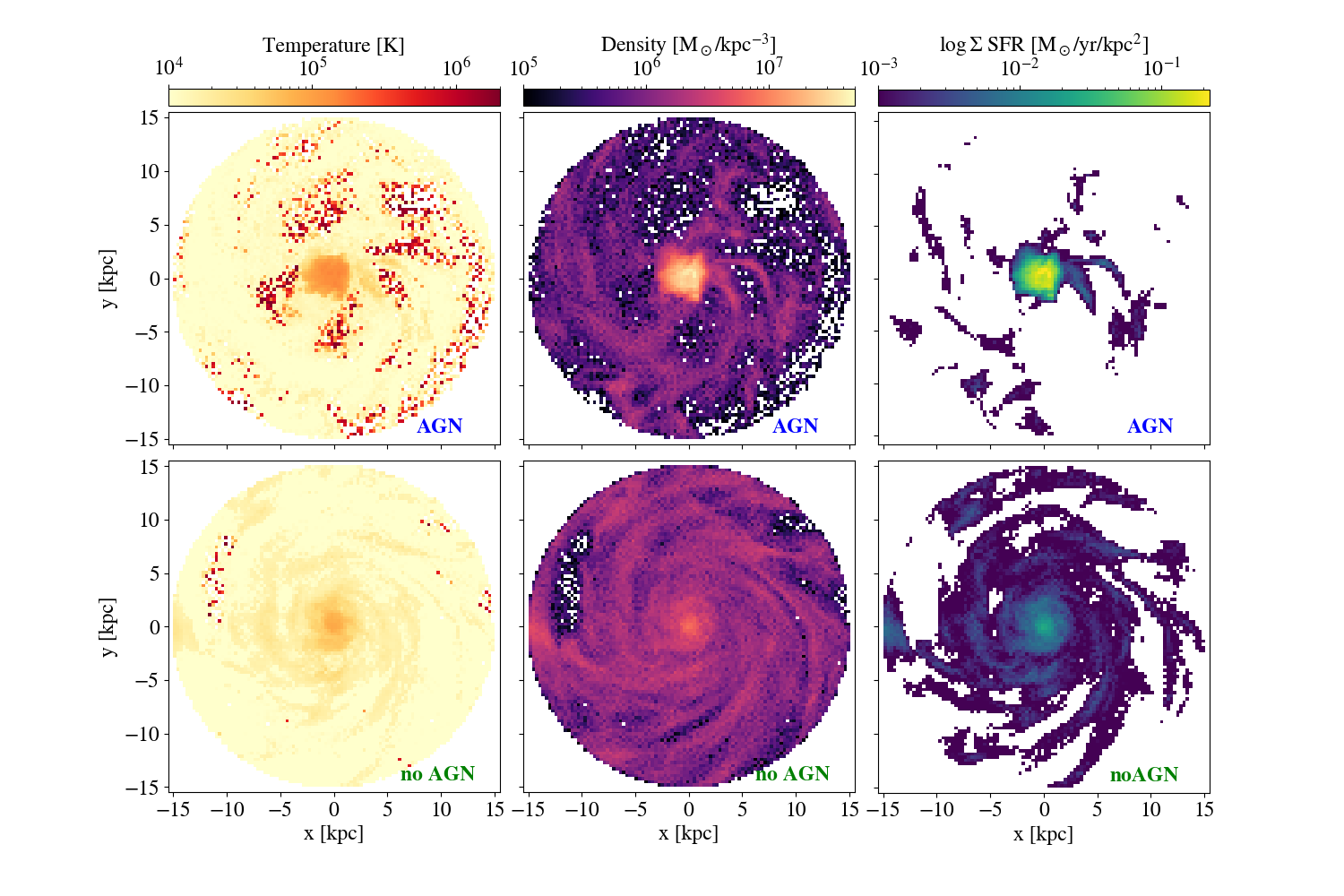}
\centering
\captionof{figure}{Face-on maps of our simulated galaxy within 15~kpc for the corresponding \textit{peak 5} in Figs. \ref{fig:outflowcheck} and \ref{fig:outflowsdiagram}. From left to right: gas temperature, gas density, and surface SFR. Top row: AGN simulation; bottom row: no-AGN counterpart.}
\label{fig:map238}

\end{minipage}
\clearpage

\noindent\begin{minipage}{\textwidth}

As noted in Section \ref{sec:outflowproperties}, we selected \textit{peak 4} for our primary analysis because it represents a clean outflow episode where the gas dynamics are dominated by outward motion. However, the absolute maximum pressure reached during the galaxy's history occurs at \textit{peak 5} (see Fig. \ref{fig:outflowcheck}). We provide here the maps and kinematic analysis for this peak to demonstrate that the underlying physics of the AGN-driven bubbles remains consistent even in the most extreme episodes.\\

Figure \ref{fig:map238} shows the spatial distribution of gas temperature, density and surface SFR at the moment of the selected outflow event. As mentioned in the main text, we exclude this peak from the primary analysis because it is partially contaminated by artificial gas inflows. These inflows are a numerical artifact of the BH repositioning scheme, which moves gas toward the BH center during rapid potential shifts.\\

Despite this numerical noise in the inflow component, Figures \ref{fig:map238} and \ref{fig:vel238} illustrate that the outflowing signatures are consistent with our findings for \textit{peak 4}. The radial velocity distribution exhibits a prominent high-velocity tail extending beyond 500 km/s, and the line-of-sight velocity profiles show the same characteristic broad wings. This confirms that the physical impact of the AGN to heat dense gas and accelerate it to high velocities is a robust feature of the model across all significant feedback events.\\

{\centering\includegraphics[scale=0.5]{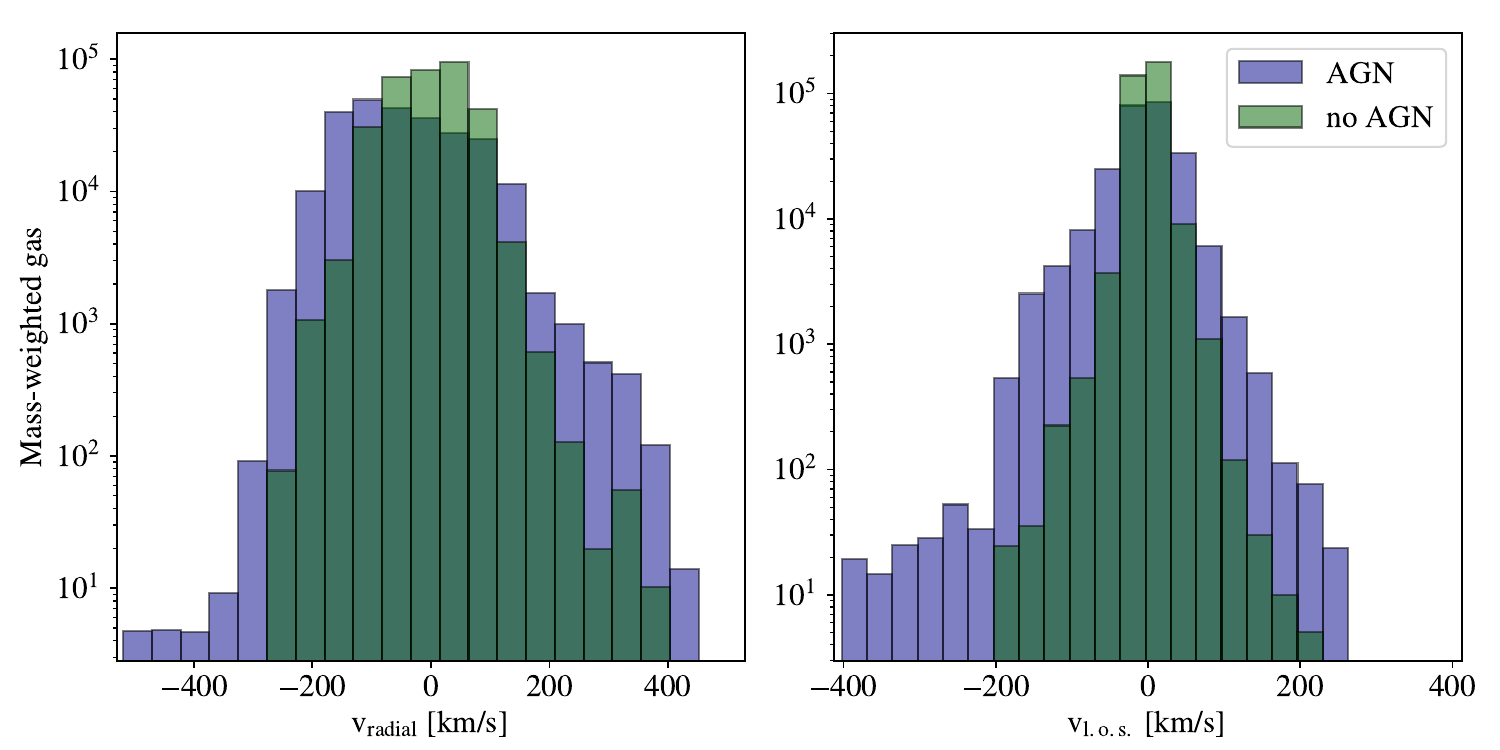}
\captionof{figure}{Radial velocity (right) and line-of-sight velocity (left) histograms of gas within 10\% of the virial radius for the AGN (blue) and no-AGN (green) simulations for the galaxy corresponding to \textit{peak 5} in Figs. \ref{fig:outflowcheck} and \ref{fig:outflowsdiagram}.}
\label{fig:vel238}}

\section{BPT evolution over different metallicity modelling}
\label{app:BPTevols}

\begin{center}
\includegraphics[width=0.5\textwidth]{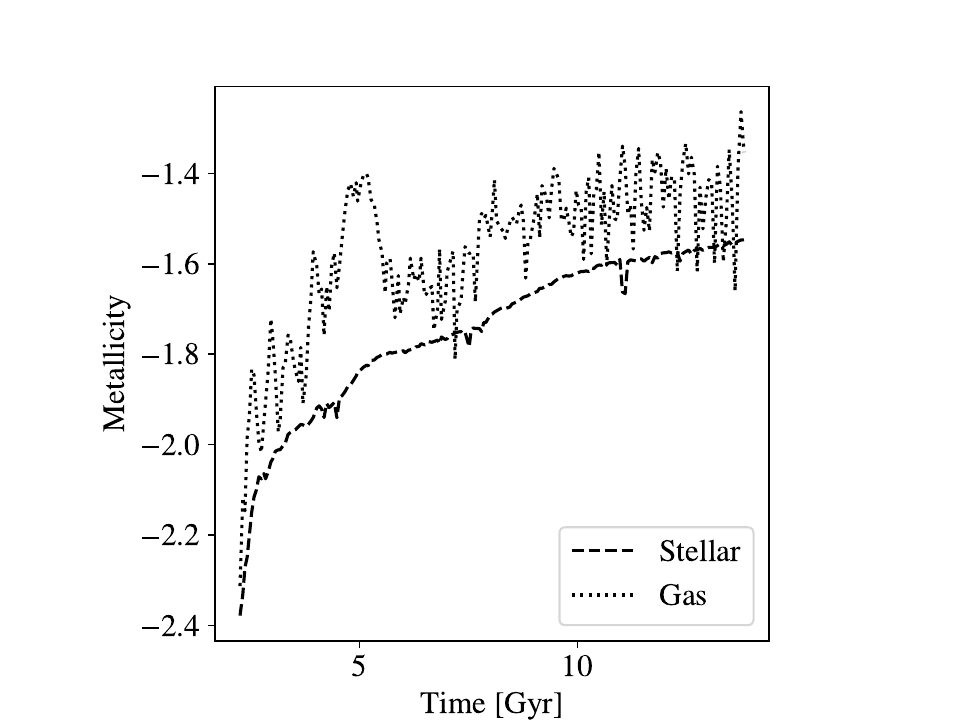}
\captionof{figure}{Mean metallicity within 1kpc over time for stellar particles and gas cells, shown with dotted and dashed lines, respectively.}
\label{fig:metals}
\end{center}

The choice of metallicity for the NLR photoionisation modelling is a key parameter that can influence the position of a galaxy in the BPT diagram. In Section \ref{sec:obscomparison}, we justified the use of the mean stellar metallicity within the central 1 kpc as a stable proxy for the chemical enrichment of the nuclear environment. Here, we present the BPT evolution using different metallicity choices.\\

\end{minipage}

\clearpage
\noindent\begin{minipage}{\textwidth}

Figure \ref{fig:metals} compares the temporal evolution of the mean stellar and gas average metallicities within the central 1 kpc. While both metallicities follow the same long-term upward trend as the galaxy matures, the gas metallicity is significantly more bursty. These fluctuations are driven by the AGN duty cycle, where episodic feedback ejects metal-rich gas from the center, while subsequent inflows of relatively metal-poor gas from the halo temporarily dilute the central reservoir. The stellar metallicity, by contrast, provides a smoother record of the long-term enrichment history and is less sensitive to these transient hydrodynamic events.\\

By comparing the BPT tracks in Fig. \ref{fig:BPTevol} with those in Fig. \ref{fig:BPTgasZ}, we observe that using the gas-phase metallicity leads to significantly higher metallicity values throughout the galaxy's history compared to the stellar proxy. This higher metal content has a clear impact on the final evolutionary stages. The higher metallicities, coupled with the decrease in the BH accretion and NLR ionisation, drive the tracks toward the LINERs region, eventually crossing the traditional AGN/LINER boundary. While the stellar-based model in Fig. \ref{fig:BPTevol} typically remains within the AGN regions, modelling the NLR using the gas metallicity places the galaxy firmly in the LINER region during its final evolutionary stages.\\

This demonstrates that while the fundamental migration from the SF region is a robust physical outcome of the BH growth, the specific classification of the galaxy at low redshift is sensitive to the enrichment level of the nuclear gas reservoir. Ultimately, the consistent shift towards higher [N\,II]/H$\alpha$ ratios across both models confirms that the overall spectroscopic life cycle is preserved, regardless of whether a smooth or more variable metallicity indicator is applied. \\

Despite these differences in the absolute classification at low redshift, the fundamental evolutionary tracks are remarkably well-preserved across all models. Whether using stellar or gas metallicity, the galaxy consistently begins its life in the SF region and migrates through the composite area toward higher excitation and enrichment states. This consistency confirms that the predicted spectroscopic life cycle of the simulated AGN is a robust physical result, independent of the specific baryonic proxy used to model the NLR emission.\\

{\centering\includegraphics[width=0.9\textwidth]{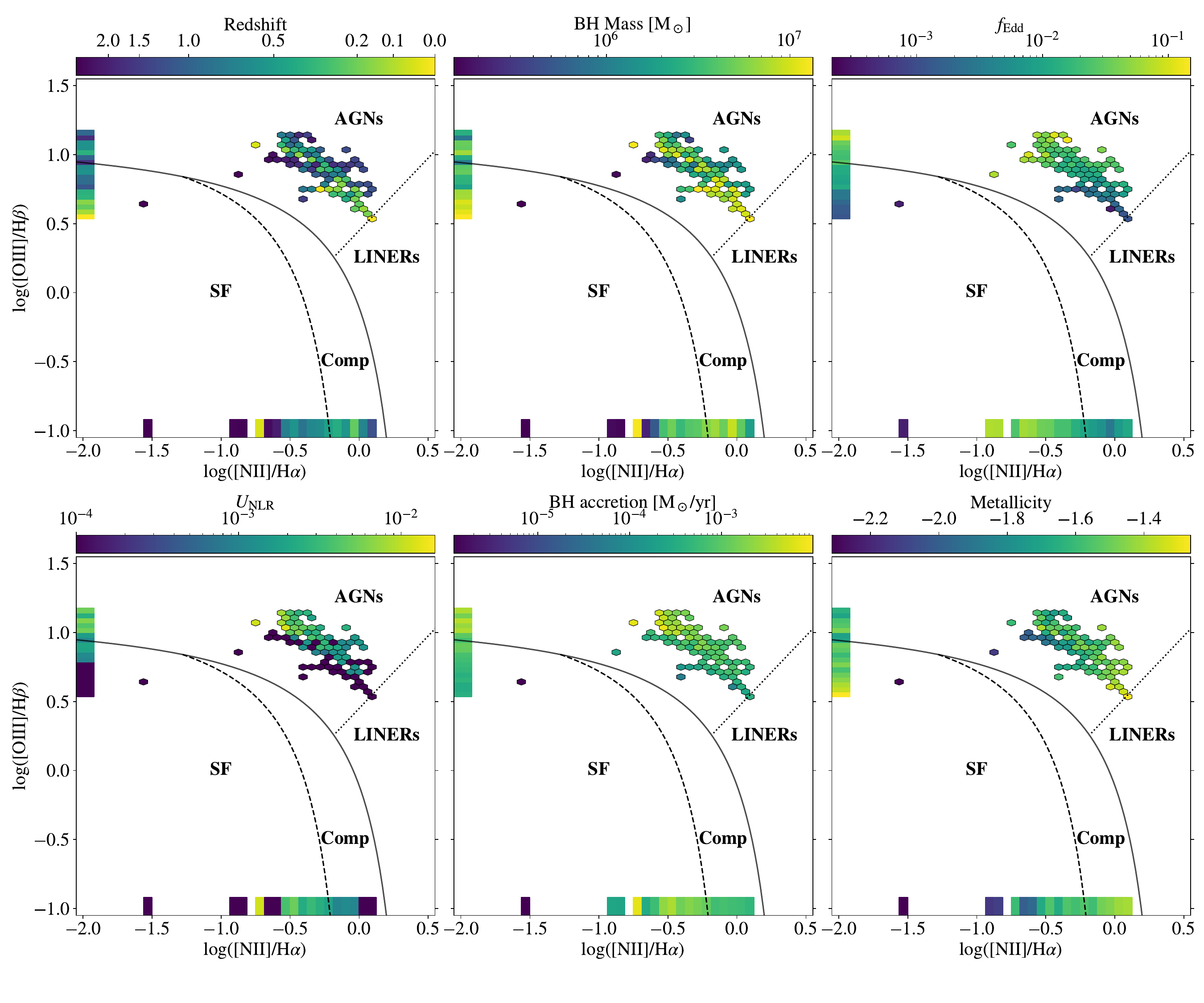}\captionof{figure}{Same as Fig. \ref{fig:BPTevol}, but using the mean gas metallicity within 1 kpc of the BH to model the NLR.}
\label{fig:BPTgasZ}}

\end{minipage}

\end{appendix}

\end{document}